\documentclass{aa}

\usepackage{graphicx}
\usepackage{xcolor}
\definecolor{laz}{RGB}{0, 150, 0}

\usepackage{natbib,twoopt}
\usepackage[breaklinks=true, colorlinks=true,linkcolor=blue, citecolor=blue]{hyperref}
\bibpunct{(}{)}{;}{a}{}{,}             
\makeatletter
  \newcommandtwoopt{\citeads}[3][][]{\href{http://adsabs.harvard.edu/abs/#3}%
    {\def\hyper@linkstart##1##2{}%
     \let\hyper@linkend\@empty\citealp[#1][#2]{#3}}}
  \newcommandtwoopt{\citepads}[3][][]{\href{http://adsabs.harvard.edu/abs/#3}%
    {\def\hyper@linkstart##1##2{}%
     \let\hyper@linkend\@empty\citep[#1][#2]{#3}}}
  \newcommandtwoopt{\citetads}[3][][]{\href{http://adsabs.harvard.edu/abs/#3}%
    {\def\hyper@linkstart##1##2{}%
     \let\hyper@linkend\@empty\citet[#1][#2]{#3}}}
  \newcommandtwoopt{\citeyearads}[3][][]%
    {\href{http://adsabs.harvard.edu/abs/#3}
    {\def\hyper@linkstart##1##2{}%
     \let\hyper@linkend\@empty\citeyear[#1][#2]{#3}}}
\makeatother
\defcitealias{das2021}{Paper~I}
\defcitealias{das2024}{Paper~II}
\usepackage{txfonts}
\begin{document}

   \title{A theoretical framework for BL~Her stars\\ III. A case study: Robust light curve optimisation in the LMC}

   \author{Susmita Das\inst{1,2,3},
           L\'aszl\'o Moln\'ar\inst{1,2,4},
        G\'abor B. Kov\'acs\inst{1,2,4,5},
        Radoslaw Smolec \inst{6},
        Meridith Joyce\inst{1,2,7},
            Shashi M. Kanbur\inst{8},  
            Tam\'as Szklen\'ar\inst{1,2},
            Anupam Bhardwaj\inst{3},
          Harinder P. Singh\inst{9},
          Marcella Marconi\inst{10}
           \and
          Vincenzo Ripepi \inst{10}
          }

   \institute{Konkoly Observatory, HUN-REN Research Centre for Astronomy and Earth Sciences, Konkoly-Thege Mikl\'os \'ut 15-17, H-1121, Budapest, Hungary\\
              \email{susmita.das@csfk.org}
        \and
             CSFK, MTA Centre of Excellence, Budapest, Konkoly Thege Miklós út 15-17., H-1121, Hungary
         \and
            Inter-University Center for Astronomy and Astrophysics (IUCAA), Post Bag 4, Ganeshkhind, Pune 411007, India
        \and
            ELTE E\"otv\"os Lor\'and University, Institute of Physics and Astronomy, 1117, P\'azm\'any P\'eter s\'et\'any 1/A, Budapest, Hungary
        \and 
            ELTE E\"{o}tv\"{o}s Lor\'{a}nd University, Gothard Astrophysical Observatory, Szombathely, Szent Imre h. u. 112., H-9700, Hungary
        \and
            Nicolaus Copernicus Astronomical Center, Polish Academy of Sciences, Bartycka 18, PL-00-716 Warsaw, Poland
        \and
            University of Wyoming, 1000 E University Ave, Laramie, WY USA
        \and
            Department of Physics, State University of New York Oswego, Oswego, NY 13126, USA
        \and
             Department of Physics \& Astrophysics, University of Delhi, Delhi 110007, India
        \and
             INAF-Osservatorio Astronomico di Capodimonte, Salita Moiariello 16, 80131, Naples, Italy        
         }
\authorrunning{S. Das et al.}
   \date{Received xxx; accepted xxx}

 
  \abstract
   {In the era of precision stellar astrophysics, classical pulsating stars play a crucial role in probing the cosmological distance scale, thanks to their period-luminosity ($PL$) relations. Therefore, it is important to constrain their stellar evolution and pulsation models not only through a comparison of empirical and theoretical $PL$ relations and properties at mean light, but also using their light curve structure over the complete pulsation cycle.}
{We carry out an extensive light curve comparison of BL~Her stars using observations from $Gaia$ DR3 and stellar pulsation models computed using \textsc{mesa-rsp} with the goal to obtain the best-matched modeled-observed pairs for BL~Her stars in the LMC.}
   {We use the Fourier decomposition technique to analyse the light curves in the $G$ band obtained from $Gaia$ DR3 and from \textsc{mesa-rsp} and 
   use a robust light curve fitting approach to score the modeled-observed pairs with respect to their pulsation periods and over their Fourier parameter space.}
   {We obtain the best-fit models for 48 BL~Her stars in the LMC and thereby provide the stellar parameter estimates of these stars, 30 of which are labelled as the \textit{gold sample} with superior light curve fits. We find a relatively flat distribution of stellar masses between 0.5--0.65\,$M_\odot$ for the \textit{gold sample} of modeled-observed pairs. An interesting result is that the majority of the best-matched models in the \textit{gold sample} are computed using the convection parameter sets without radiative cooling. The period-Wesenheit relation for the best-matched \textit{gold sample} of 30 BL~Her models exhibits a slope of $-2.805 \pm 0.164$ while the corresponding period-radius relation exhibits a slope of $0.565 \pm 0.035$, both in good agreement with the empirical $PW$ and $PR$ slopes from BL~Her stars in the LMC, respectively. We also used the Wesenheit magnitudes of the 30 best-matched modeled-observed pairs to estimate a distance modulus of $\mu_{\rm LMC} = 18.582 \pm 0.067$ to the LMC, which lies within the bounds of previous literature values. We also discuss the degeneracy in the stellar parameters of the BL~Her models that result in similar pulsation periods and light curve structure, and highlight that caution must be exercised while using the stellar parameter estimates.}
   {}


   \maketitle


\section{Introduction}

BL~Herculis (BL~Her) stars are the shortest-period type~II Cepheids (T2Cs) with pulsation periods between 1 and 4 days \citep{soszynski2018}. Their evolutionary status is rather complex; these stars are predicted to be double-shell burning while still approaching their Asymptotic Giant Branch (AGB) track \citep{bono2020}. Similar to the other classical pulsators, BL~Her stars also obey well-defined period-luminosity ($PL$) relations but with the added advantage of exhibiting weak or negligible effects of  metallicity over multiple wavelengths, as demonstrated by empirical \citep{matsunaga2006, matsunaga2009, matsunaga2011, groenewegen2017b,bhardwaj2022,ngeow2022} and theoretical \citep{criscienzo2007, das2021, das2024} studies. BL~Her stars used in combination with RR~Lyraes and the tip of the red giant branch (TRGB) may therefore prove useful as an alternative to classical Cepheids in the calibration of the extragalactic distance scale \citep{majaess2010,beaton2016,braga2020,das2021}.

The characterization of the light curve structure of classical pulsators using Fourier analysis dates back to the pioneering works of \citet{simon1981} and \citet{simon1982} for Cepheids and RR~Lyrae stars, respectively, and has since been used to study the progressions of the Fourier coefficients as a function of pulsation periods. The earliest comparisons of theoretical and observed light curves using Fourier decomposition were carried out by \citet{simon1983} for classical Cepheids and by \citet{feuchtinger1997} and \citet{kovacs1998} for RR~Lyrae stars. A comparison of the theoretical and observed light curves of classical pulsators not only provides stringent constraints for the pulsation models but also helps in understanding the theory of stellar pulsation. In particular, it has been useful to study the phenomenon of the existence of bump Cepheids and the Hertzsprung progression in classical Cepheids \citep{buchler1990, bono2002, keller2002, keller2006, marconi2024} and period doubling in BL~Her stars \citep{buchler1992, soszynski2011, smolec2012a}.

Light curve structures are affected by the global stellar parameters as well as the internal physics of the stars; the existence of the correlation among the Fourier parameter $\phi_{31} \equiv \phi_3 - 3\phi_1$, pulsation period and metallicity of RR~Lyrae stars \citep{jurcsik1996, smolec2005, nemec2013} is one of the most notable results in this regard. Stellar pulsation models are therefore useful in estimating the global physical parameters of the stars. \citet{wood1997} were the first to use the light curve model fitting technique for a classical Cepheid whereas \citet{bono2000d} were the first to apply the technique to an RR~Lyrae pulsator. Subsequently, there have been several studies with fitting light curves and also radial velocity curves, when available, for Cepheids \citep{bono2002, keller2002, keller2006, marconi2013a, marconi2017a, bhardwaj2017a} as well as RR~Lyrae stars \citep{difabrizio2002, marconi2005, marconi2007a, marconi2007b, das2018}. \citet{marconi2013b} also estimated the physical parameters for an eclipsing binary Cepheid using a grid of models that matched the observed light curves, radial velocity curves and radius curves. More recently, \citet{bellinger2020} and \citet{kumar2024} have trained artificial neural networks on stellar pulsation models to predict the global stellar parameters based on their pulsation periods and light curve structure of classical pulsators.

This paper is the third in the series, subsequent to
\citet{das2021} and \citet{das2024} (hereafter \citetalias{das2021, das2024}) which uses a very fine grid of convective BL~Her models computed using the state-of-the-art 1D non-linear Radial Stellar Pulsation (\textsc{rsp}) tool within the \emph{Modules for Experiments in Stellar Astrophysics} \citep[\textsc{mesa},][]{paxton2011,paxton2013,paxton2015,paxton2018,paxton2019,jermyn2023} software suite and encompassing a wide range of input parameters: metallicity ($-2.0\; \mathrm{dex} \leq \mathrm{[Fe/H]} \leq 0.0\; \mathrm{dex}$), stellar mass (0.5--0.8\,$M_{\odot}$), stellar luminosity (50--300\,$L_{\odot}$), and effective temperature (full extent of the instability strip in steps of 50\,K; 5900--7200\,K for 50\,$L_{\odot}$ and 4700--6550\,K for 300\,$L_{\odot}$). In \citetalias{das2021} and \citetalias{das2024}, we derived theoretical period-radius ($PR$) and $PL$ relations for BL~Her models in the Johnson-Cousins-Glass bands ($UBVRIJHKLL'M$) and in the $Gaia$ passbands ($G$, $G_{BP}$, and $G_{RP}$), respectively, and thereby compared the theoretical relations at mean light with the empirical relations from \citet{matsunaga2006, matsunaga2009, matsunaga2011, bhardwaj2017c, groenewegen2017b, ripepi2023}. We also found negligible effects of metallicity on the theoretical $PL$ relations in wavelengths longer than $B$ band, consistent with empirical results. 

Although the empirical $PL$ relations from the BL~Her stars show statistically similar slopes with those from the models at mean light, it is important to remember that relations at mean light exhibit an averaged behaviour over the entire pulsation cycle. \citetalias{das2024} also showed that the light curves of the BL~Her models are affected by the choice of convection parameters. In the percent-level precision era, it is crucial to compare not just the empirical and theoretical relations at mean light but also their light curve structure over the complete pulsation cycle. In this paper, we therefore carry out an extensive light curve analysis comparing the theoretical and observed $G$ band light curves of BL~Her stars in the Large Magellanic Cloud (LMC) in an effort to provide better constraints on stellar pulsation theory. The structure of the paper is as follows: the theoretical and empirical data used in this analysis are briefly described in Section~\ref{sec:data}, followed by a brief description of the light curve parameters obtained using the Fourier decomposition technique in Section~\ref{sec:lc}. The best--matched observed--model pairs for the BL~Her stars in the LMC and the estimated stellar parameters for these stars are presented in Section~\ref{sec:model-obs} and validated in Section~\ref{sec:ogle}. The LMC distance estimated using the best--matched observed--model pairs is provided in Section~\ref{sec:distance}. We also investigate the degeneracy in the stellar parameters of the BL~Her models resulting in similar pulsation periods and light curve structure and corresponding to the same star in Section~\ref{sec:degeneracy}. Finally, the results of the study are summarized in Section~\ref{sec:results}.

\section{The Data}
\label{sec:data}

\subsection{Stellar pulsation models}

We use the non-linear radial stellar pulsation models that were computed using \textsc{mesa-rsp} \citep{smolec2008,paxton2019} in \citetalias{das2021} with input parameters typical for BL~Her stars and their corresponding $Gaia$ passband light curves from \citetalias{das2024}. The input parameters that go into constructing a stellar pulsation model are its chemical composition -- hydrogen mass fraction $X$ and heavy element mass fraction $Z$, stellar mass $M$, stellar luminosity $L$, and effective temperature $T_{\rm{eff}}$. The interested reader is referred to \citetalias{das2021}, \citetalias{das2024} for a detailed description of the BL~Her models computed and used in this analysis. The grid of models was computed for seven metal abundances, $Z = 0.00014, 0.00043, 0.00061, 0.00135, 0.00424, 0.00834, 0.013$ (see Table~\ref{tab:composition} for more details), for stellar masses in the range $M/M_{\odot} \in \left\langle 0.5, 0.8\right\rangle$ with a step of $0.05 M_{\odot}$, stellar luminosities $L/L_{\odot} \in \left\langle 50, 300\right\rangle$\footnote{The low mass models $M/M_{\odot} \in [0.5, 0.6]$ were computed with stellar luminosity $L/L_{\odot} \in \left\langle 50, 200\right\rangle$ only.} with a step of $50 L_{\odot}$, and effective temperatures $T_{\rm{eff}} \in \left\langle4000, 8000\right\rangle$ K with a step of 50 K. Each grid with the above-mentioned $ZXMLT_{\rm eff}$ input stellar parameters was computed with four sets of convection parameters \citep[see, Table~4 of][]{paxton2019}: set~A (the simplest convection model), set~B (with radiative cooling), set~C (with turbulent pressure and turbulent flux), and set~D (with all the effects added simultaneously). The non-linear computations were carried for 4000 pulsation cycles each. Only those models with non-linear pulsation periods between 1--4 days were considered to be BL~Her models \citep{soszynski2018} followed by the check of single periodicity and full-amplitude stable pulsation conditions of the models\footnote{The fractional growth of the kinetic energy per pulsation period $\Gamma$ does not vary by more than 0.01, the amplitude of radius variation $\Delta R$ by more than 0.01~$R_{\odot}$, and the pulsation period $P$ by more than 0.01~d over the last 100 cycles of the completed integration (see Fig.~2 of \citetalias{das2021} for a detailed pictorial representation).}. The total numbers of BL~Her models accepted for subsequent analysis are as follows: 3266~(set~A), 2260~(set~B), 2632~(set~C) and 2122~(set~D).

As mentioned in \citetalias{das2021} and \citetalias{das2024}, our grid of models include stellar masses higher than what is predicted for BL~Her stars which are considered to be low-mass stars evolving from the blue edge of the Zero Age Horizontal Branch (ZAHB) towards the AGB \citep{gingold1985, bono1997a, caputo1998}. In addition, the higher stellar mass (M\,$> 0.6\,M_{\odot}$) and lower metallicity ($Z=0.00014$) models may be considered typical of ZAHB or evolved RR~Lyrae stars. The distinction between BL~Her and RR~Lyrae stars can be complicated: \citet{braga2020}, for example, found an overlap between the two types above 1.0~d period in $\omega$~Cen. However, they found that evolved RR~Lyrae stars separate in amplitudes and in certain Fourier terms. Concurrently, \citet{netzel2023} found that seismic masses for overtone RR~Lyrae stars extend downwards to 0.5--0.6 $M_\odot$, i.e., into the BL~Her range. While we use the complete set of BL~Her models for a comparison with the observed stars in this study, we give preference to the lower stellar mass models ($M \leq 0.65\,M_{\odot}$) as a better fit to the observed stars, wherever possible, following the work of \citet{marconi2007a} where 0.65 $M_{\odot}$ was used as the upper limit for BL~Her models.

\subsection{Observations from $Gaia$ DR3}

\begin{figure*}
\centering
\includegraphics[scale = 1]{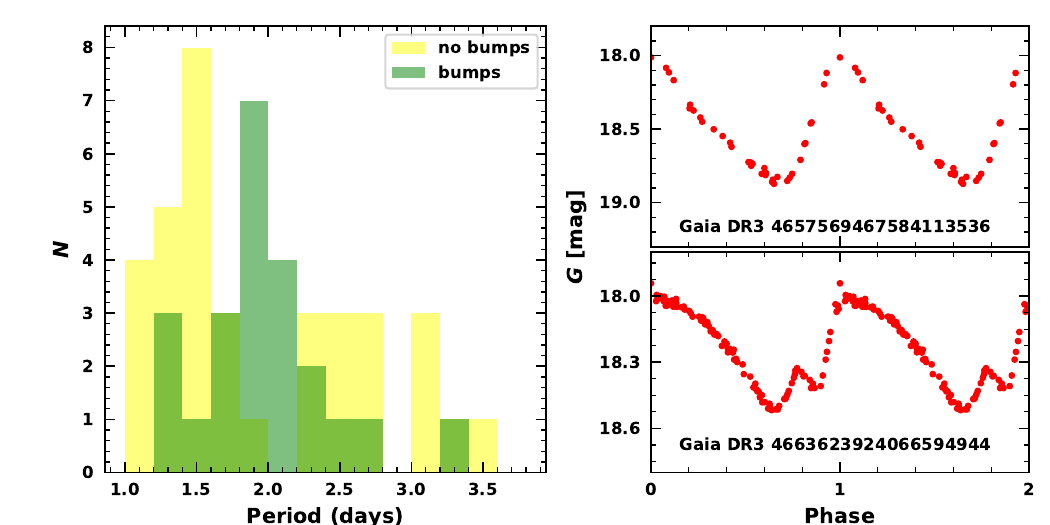}
\caption{Period distribution of the observed BL~Her stars in the LMC with and without the presence of the bump feature in their $G$ band light curve structure. Examples of BL~Her stars exhibiting no bump or bump in their light curves are displayed in the upper right and lower right panels respectively.}
\label{fig:period_histogram}
\end{figure*}

We use the photometric data of 58 BL~Her stars in the LMC \citepalias[for more details, see Table~3 of][]{das2024} in the three $Gaia$ passbands ($G$, $G_{BP}$, and $G_{RP}$), available from the European Space Agency's (ESA) $Gaia$ mission \citep[$Gaia$ DR3;][]{prusti2016, vallenari2023}. The detailed light curve comparison in this analysis is carried out using the $G$ band light curves only, since the empirical light curves contain better photometric data with a good phase coverage as compared to those in the $G_{BP}$ and $G_{RP}$ passbands. However, we also make use of the \textit{mean} $GG_{BP}G_{RP}$ magnitudes later in the analysis to obtain the Wesenheit magnitudes, as discussed in Section~\ref{sec:distance}. We visually separate out the BL~Her stars that exhibit a bump feature in their $G$ band light curves from those that do not; example light curves of BL~Her stars with and without the bump feature in their light curve structure are displayed in Fig.~\ref{fig:period_histogram}. We also study the period distribution of BL~Her stars in the LMC with and without the presence of the bump feature in their $G$ band light curve structure in Fig.~\ref{fig:period_histogram}. We find that the distributions are different: BL~Her stars with bump feature show a maximum at around 2 days, while those without bump peak at shorter periods. While the presence and the progression of the bump feature in the light curves of BL~Her stars is known and is attributed to a phenomenon analogous to the Hertzsprung progression for classical Cepheids with periods around 10 days \citep{petersen1980, buchler1992, moskalik1993, marconi2007a}, it remains to be investigated in details. 23 BL~Her stars in the LMC exhibit a bump in their $G$ band light curves while the remaining 35 BL~Her stars do not.

\section{A description of the light curve parameters}
\label{sec:lc}
We study the light curve structure of the BL~Her stars using the Fourier decomposition technique. The theoretical and observed light curves in the $Gaia$ passbands are fitted with the Fourier sine series \citep[see, for example,][]{deb2009, das2020} of the form:
\begin{equation}
m(x) = m_0 + \sum_{k=1}^{N}A_k \sin(2 \pi kx+\phi_k),
\label{eq:fourier}
\end{equation}

\noindent where $x$ is the pulsation phase, $m_0$ is the mean magnitude, and $N$ is the order of the fit. The theoretical light curves are fitted with $N=20$. For observed light curves, an optimum value of $N$ is important so as not to fit the noise with an $N$ too large or include systematic deviations with an $N$ too small. To this end, we use the Baart’s condition \citep{baart1982} and vary $N$ from 4 to 8 to obtain the optimum order of fit. However, if we have a good phase coverage, we can use a reasonably higher order of fit ($N \leq 8$) without running into the risk of over-fitting. This allows us to obtain higher order Fourier parameters (described below) which in turn helps us in constraining modeled-observed pairs better. The 35 BL~Her stars that do not exhibit bumps are fitted with the order of fit $N$ as obtained from the Baart’s condition. Of the 23 BL~Her stars that exhibit bumps, 13 BL~Her stars have a good phase coverage and are fitted with $N=7$ while the rest 10 are fitted with the optimum order of fit using the Baart’s condition.

\subsection{Fourier parameters}

The Fourier amplitude and phase coefficients ($A_k$ and $\phi_k$) are used to obtain the Fourier amplitude and phase parameters, respectively as follows:
\begin{equation}
\begin{aligned}
R_{k1} &= \frac{A_k}{A_1}, \\
\phi_{k1} &= \phi_k - k\phi_1,
\label{eq:params}
\end{aligned}
\end{equation}
\noindent where $k > 1$ and $ 0 \leq \phi_{k1} \leq 2\pi$. 

\subsection{Amplitude}

The peak-to-peak amplitude $A$ of the light curve is defined as the difference in the minimum and the maximum of the light variations:
\begin{equation}
    A_{\lambda} = (M_{\lambda})_{\rm min} - (M_{\lambda})_{\rm max},
\end{equation}
where $(M_{\lambda})_{\rm min}$ and $(M_{\lambda})_{\rm max}$ are the minimum and maximum magnitudes obtained from the Fourier fits in the $\lambda$ band, respectively. 

\subsection{Skewness and acuteness}
We also use a robust definition for the skewness of light curve which is unaffected by the presence of bumps. The definition of skewness is as follows \citep{elgar1987}:
\begin{equation}
    S_k = \frac{\left\langle\mathcal{H} (m)^3 \right\rangle}{\left\langle m \right\rangle^{3/2}},
\end{equation}
where $m$ is the mean-reduced light curve of the star, the brackets denote the mean over one pulsation period\footnote{In general, this should be for a large number of periods, but since we work with the Fourier series, there is no cycle-to-cycle variation.}. $\mathcal{H}$ is the Hilbert transformation \citep[Chapter 3]{duoandikoetxea2001}.

Acuteness is defined as the ratio of the phase duration of magnitude fainter than the mean magnitude to that of magnitude brighter than the mean magnitude \citep{stellingwerf1987}:
\begin{equation}
A_c = \frac{1}{\phi_{fw}}-1,
\label{eq:Ac}
\end{equation}
where $\phi_{fw}$ is the phase duration of the magnitude which is brighter than the mean magnitude (equal to the full-width at half-maximum of the light curve).

\section{Estimating the stellar parameters of BL~Her stars in the LMC}
\label{sec:model-obs}

\begin{figure*}
\centering
\includegraphics[scale = 1]{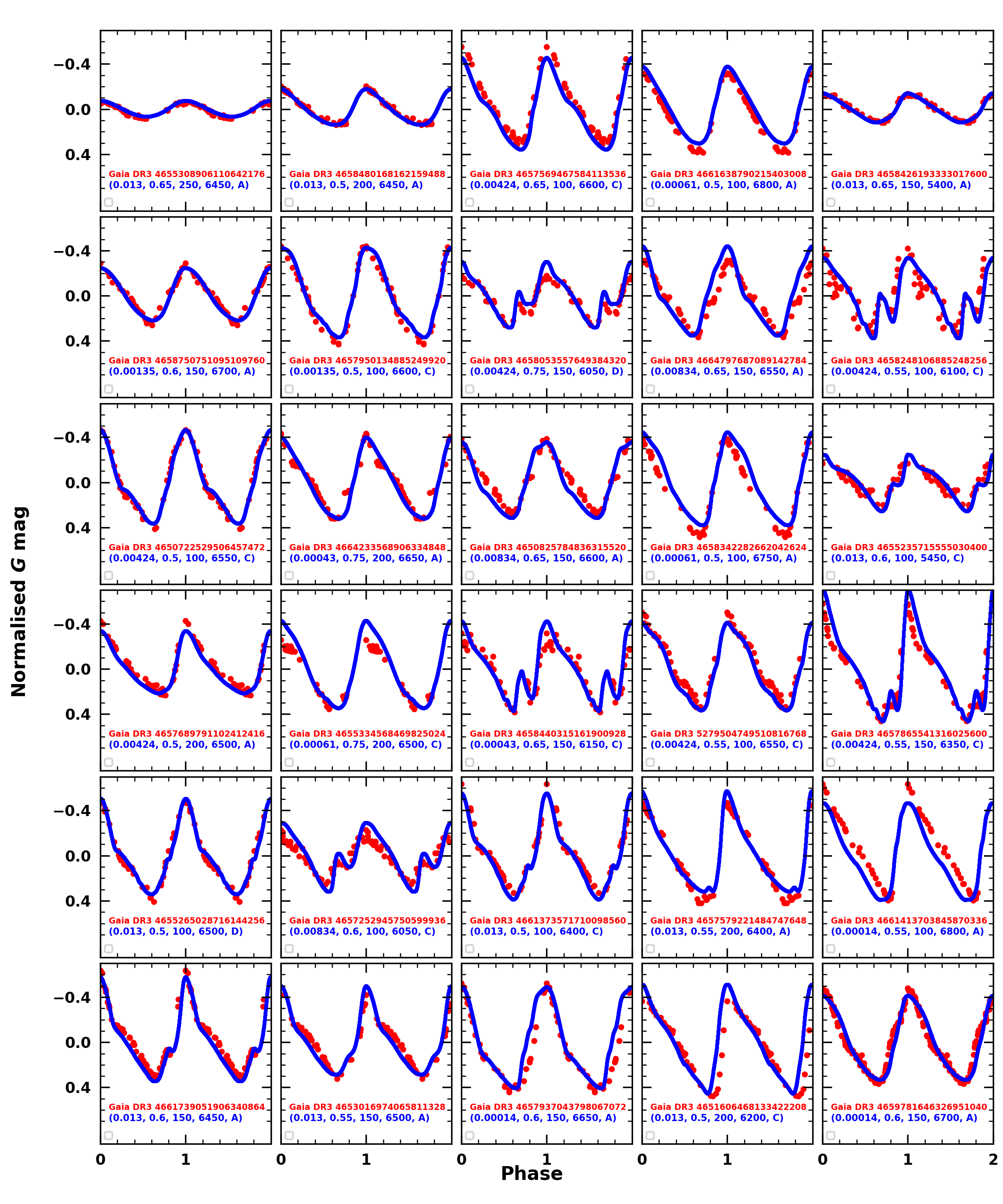}
\caption{The light curves (normalized with respect to their mean magnitudes) of the 30 BL~Her stars in the LMC (in red) with their best-matched models (in blue). These 30 accepted modeled-observed pairs are considered as the \textit{gold sample} with superior model fitting. The input stellar parameters of the corresponding models are included in the format ($Z, M/M_{\odot}, L/L_{\odot}, T_{\rm eff}$, convection set) in each sub-plot.}
\label{fig:LC_gold}
\end{figure*}

\begin{figure*}
\centering
\includegraphics[scale = 1]{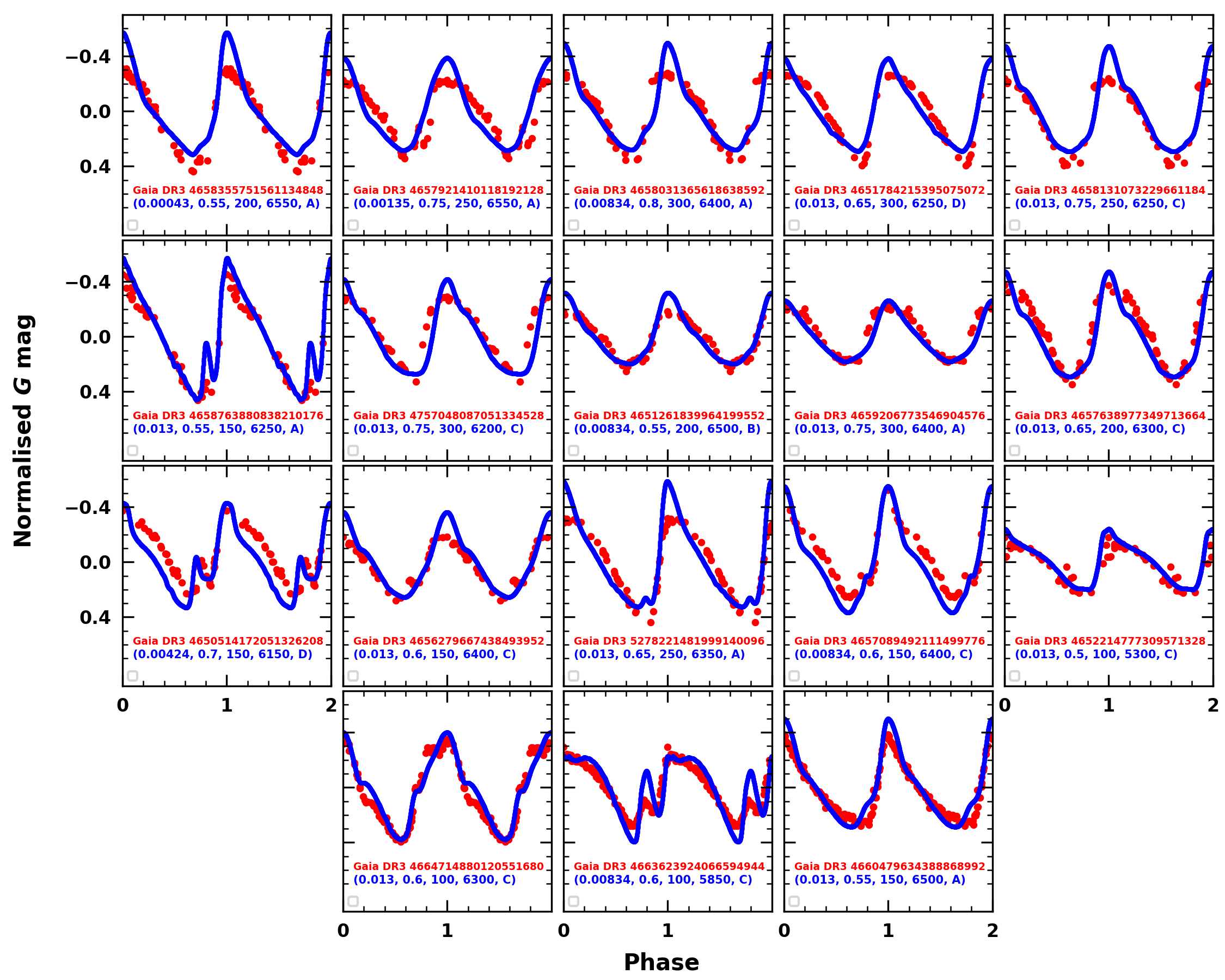}
\caption{Same as Fig.~\ref{fig:LC_gold} but for the modeled-observed pairs considered as the \textit{silver sample}.}
\label{fig:LC_silver}
\end{figure*}

 We now compare the theoretical and observed $G$ band light curve structure of the BL~Her stars in the LMC that have similar pulsation periods. Note that we carry out the comparison for all observed BL~Her stars in the LMC with the complete set of BL~Her models across the four sets of convection parameters. This allows the possibility for stars with different effective temperatures to prefer models with different convection parameters, following the work of \citet{kovacs2023} where they calibrated $T_{\rm eff}$-dependent convective parameters for RRab stars in M3. 

\subsection{BL~Her stars with no bumps}

Higher-order Fourier amplitude parameters capture the bump feature in light curves. Therefore, to find the best-fit matches for stars that do not exhibit bumps in their light curves, we restrict our model sample to stars where the higher order Fourier amplitude parameters do not exceed:
\begin{equation}
R_{k1} \leq 0.015, 
\end{equation}
where $8\leq k \leq 11$. Limiting the higher order Fourier amplitude parameters rejects the BL~Her models with bumps while still retaining BL~Her models that can be matched with BL~Her stars exhibiting the characteristic saw-tooth light curves without bumps.

Next, we constrain the sample of models with pulsation periods and amplitudes similar to those from observed BL~Her stars in the LMC using the condition:
\begin{equation}
\begin{aligned}
|\log(P)_{\rm mod} - \log(P)_{\rm obs}| \leq 0.01, \\
|A_{\rm mod} - A_{\rm obs}| \leq 0.2 {\, \rm mag},
\end{aligned}
\label{eq:periodconstraint}
\end{equation}
where $P_{\rm mod}$ and $P_{\rm obs}$ are the theoretical and the observed pulsation periods, respectively and $A_{\rm mod}$ and $A_{\rm obs}$ are the theoretical and the observed peak-to-peak amplitudes in the $G$ band, respectively.

For a rigorous comparison, it is useful to construct a mathematically robust definition of the goodness-of-fit for each modeled-observed pair \citep[see, for example,][]{smolec2013, joyce2023}. We define a parameter $d$ to compute the offsets for each modeled-observed pair $i$ with respect to their pulsation periods and over their Fourier parameter space as follows \citep[see ][]{smolec2013}:
\begin{equation}
d = \sqrt{\sum_i n_p (p_{i,\rm mod}-p_{i,\rm obs})^2/p_{i,\rm obs}^2},
\label{eq:d}
\end{equation}
where $p \in \{ \log(P), A, S_k, A_c, R_{21}, R_{31}, \phi_{21}, \phi_{31} \}$ for the theoretical ($p_{i,\rm mod}$) and observed ($p_{i,\rm obs}$) light curves, with $n_p$ as the weights for each parameter. While the rest of the parameters have equal weights of $n_p=1$, the amplitude ($A$) and the skewness ($S_k$) of the light curve  need a considerably higher weight of $n_p=10$ to ensure that we obtain the best matches between models and observations. 

\citet{smolec2013} had equal weights of $n_p=1$ for all light curve parameters involved in the determination of the parameter $d$. The reason for including a higher weight of $n_p=10$ on the amplitude and skewness parameters in this work follows from the results obtained after carrying out a few exercises with different weights on $A$ and $S_k$ (see Fig.~\ref{fig:diffN} for more) where we found that a higher weight of $n_p=10$ on the $A$ and $S_k$ parameters results in better matches of modeled-observed pairs as compared to using $n_p=1$ or $n_p=5$. Increasing the weight on the $A$ and $S_k$ parameters to $n_p=20$ does not 
improve the modeled-observed pairs further. In addition, we also tested a higher weight of $n_p=10$ on the pulsation period $\log(P)$ and found that it results in the same matches as with $n_p=1$; this is likely because we had already placed a strong constraint on the pulsation period, $|\log(P)_{\rm mod} - \log(P)_{\rm obs}| \leq 0.01$, to begin with. We would like to highlight here that while the light curve optimisation method adopted in this work is quite robust and we encourage readers to adopt this approach for similar work, the weights $n_p$ on the light curve parameters should be decided after testing what works best for a particular dataset.

\begin{table*}
\caption{BL~Her stars in the LMC with their corresponding best-matched \textsc{mesa-rsp} models with respect to their pulsation periods and light curve structure. The first four columns are from $Gaia$~DR3 observations. The absence or presence of bumps in their $G$ band light curves are denoted by ``NB'' (no bumps) or ``B'' (bumps), respectively. The Wesenheit magnitudes from observations ($W_{obs}$) and models ($W_{th}$) are used to estimate the distance modulus $\mu$ to the LMC. The next seven columns are from the corresponding best-matched models and provide stellar parameter estimations (chemical composition $ZX$, stellar mass $M$, stellar luminosity $L$, stellar radius $R$, effective temperature $T_{\rm eff}$ and the convection parameter set used). The last column is the Kolmogorov-Smirnov goodness-of-fit score for each modeled-observed pair.}
\small
\centering
\scalebox{0.82}{
\begin{tabular}{c c c c c c c c c c c c c c}
\hline\hline
$Gaia$ Source ID & Bump & $\log(P)$ & $W_{obs}$ & $W_{th}$ & $\mu$ & $Z^\dagger$ & $X$ &  $\frac{M}{M_{\odot}}$ & $\frac{L}{L_{\odot}}$ & $\log\frac{R}{R_{\odot}}$ & $T_{\rm{eff}}$ & Set & Score\\
[0.5ex]
\hline
\multicolumn{14}{c}{Gold sample}\\
\hline
4655308906110642176 & NB& 0.41 & 16.262$\pm$0.014 & -2.436$\pm$0.0 & 18.698$\pm$0.014 & 0.013 & 0.71847 & 0.65 & 250 & 1.107 & 6450 & A & 3.98e-06\\
4658480168162159488 & NB& 0.404 & 15.041$\pm$0.055 & -2.192$\pm$0.0 & 17.233$\pm$0.055 & 0.013 & 0.71847 & 0.5 & 200 & 1.06 & 6450 & A & 6.18e-07\\
4657569467584113536 & NB& 0.021 & 17.267$\pm$0.136 & -1.316$\pm$0.0 & 18.583$\pm$0.136 & 0.00424 & 0.74073 & 0.65 & 100 & 0.894 & 6600 & C & 2.17e-07\\
4661638790215403008 & NB& 0.037 & 17.617$\pm$0.127 & -1.137$\pm$0.0 & 18.754$\pm$0.127 & 0.00061 & 0.74996 & 0.5 & 100 & 0.864 & 6800 & A & 7.77e-08\\
4658426193333017600 & NB& 0.523 & 16.127$\pm$0.019 & -2.516$\pm$0.0 & 18.643$\pm$0.019 & 0.013 & 0.71847 & 0.65 & 150 & 1.151 & 5400 & A & 2.05e-08\\
4658750751095109760 & NB& 0.17 & 17.079$\pm$0.05 & -1.643$\pm$0.0 & 18.722$\pm$0.05 & 0.00135 & 0.74806 & 0.6 & 150 & 0.963 & 6700 & A & 1.17e-08\\
4657950134885249920 & NB& 0.102 & 17.266$\pm$0.171 & -1.274$\pm$0.0 & 18.54$\pm$0.171 & 0.00135 & 0.74806 & 0.5 & 100 & 0.897 & 6600 & C & 2.83e-09\\
4658053557649384320 & B& 0.265 & 16.845$\pm$0.033 & -2.081$\pm$0.0 & 18.926$\pm$0.033 & 0.00424 & 0.74073 & 0.75 & 150 & 1.06 & 6050 & D & 2.63e-09\\
4664797687089142784 & NB& 0.196 & 16.961$\pm$0.048 & -1.805$\pm$0.0 & 18.766$\pm$0.048 & 0.00834 & 0.73032 & 0.65 & 150 & 0.988 & 6550 & A & 2.61e-09\\
4658248106885248256 & B& 0.207 & 17.033$\pm$0.066 & -1.624$\pm$0.0 & 18.657$\pm$0.066 & 0.00424 & 0.74073 & 0.55 & 100 & 0.976 & 6100 & C & 2.05e-09\\
4650722529506457472 & NB& 0.119 & 17.202$\pm$0.046 & -1.345$\pm$0.0 & 18.547$\pm$0.046 & 0.00424 & 0.74073 & 0.5 & 100 & 0.903 & 6550 & C & 1.78e-09\\
4664233568906334848 & NB& 0.204 & 17.039$\pm$0.074 & -1.955$\pm$0.0 & 18.994$\pm$0.074 & 0.00043 & 0.75041 & 0.75 & 200 & 1.035 & 6650 & A & 4.8e-10\\
4650825784836315520 & NB& 0.169 & 16.905$\pm$0.068 & -1.776$\pm$0.0 & 18.681$\pm$0.068 & 0.00834 & 0.73032 & 0.65 & 150 & 0.98 & 6600 & A & 3.59e-10\\
4658342282662042624 & NB& 0.062 & 17.122$\pm$0.153 & -1.166$\pm$0.0 & 18.288$\pm$0.153 & 0.00061 & 0.74996 & 0.5 & 100 & 0.875 & 6750 & A & 3.43e-10\\
4655235715555030400 & B& 0.394 & 16.415$\pm$0.048 & -2.114$\pm$0.0 & 18.529$\pm$0.048 & 0.013 & 0.71847 & 0.6 & 100 & 1.076 & 5450 & C & 2.53e-10\\
4657689791102412416 & NB& 0.378 & 15.4$\pm$0.106 & -2.099$\pm$0.0 & 17.499$\pm$0.106 & 0.00424 & 0.74073 & 0.5 & 200 & 1.056 & 6500 & A & 1.2e-10\\
4655334568469825024 & NB& 0.242 & 16.97$\pm$0.197 & -2.054$\pm$0.0 & 19.024$\pm$0.197 & 0.00061 & 0.74996 & 0.75 & 200 & 1.057 & 6500 & C & 5.54e-11\\
4658440315161900928 &B	&0.287	&16.524$\pm$0.062	&-1.950$\pm$0.0	&18.474$\pm$0.062	&0.00043	&0.75041	&0.65	&150	&1.059	&6150	&C	&2.43e-11\\
5279504749510816768 & NB& 0.088 & 17.292$\pm$0.127 & -1.347$\pm$0.0 & 18.639$\pm$0.127 & 0.00424 & 0.74073 & 0.55 & 100 & 0.904 & 6550 & C & 1.88e-11\\
4657865541316025600 & B& 0.289 & 16.856$\pm$0.184 & -1.908$\pm$0.0 & 18.764$\pm$0.184 & 0.00424 & 0.74073 & 0.55 & 150 & 1.037 & 6350 & C & 2.4e-12\\
4655265028716144256 & NB& 0.14 & 17.424$\pm$0.146 & -1.435$\pm$0.0 & 18.859$\pm$0.146 & 0.013 & 0.71847 & 0.5 & 100 & 0.91 & 6500 & D & 1.43e-12\\
4657252945750599936 & B& 0.192 & 16.946$\pm$0.036 & -1.688$\pm$0.0 & 18.634$\pm$0.036 & 0.00834 & 0.73032 & 0.6 & 100 & 0.977 & 6050 & C & 8.38e-13\\
4661373571710098560 & NB& 0.16 & 17.069$\pm$0.076 & -1.498$\pm$0.0 & 18.567$\pm$0.076 & 0.013 & 0.71847 & 0.5 & 100 & 0.926 & 6400 & C & 3.52e-13\\
4657579221484747648 & NB& 0.381 & 16.422$\pm$0.139 & -2.23$\pm$0.0 & 18.652$\pm$0.139 & 0.013 & 0.71847 & 0.55 & 200 & 1.078 & 6400 & A & 1.92e-13\\
4661413703845870336 & NB& 0.018 & 17.217$\pm$0.085 & -1.116$\pm$0.0 & 18.333$\pm$0.085 & 0.00014 & 0.75115 & 0.55 & 100 & 0.868 & 6800 & A & 6.55e-14\\
4661739051906340864 & NB& 0.246 & 16.959$\pm$0.056 & -1.892$\pm$0.0 & 18.851$\pm$0.056 & 0.013 & 0.71847 & 0.6 & 150 & 1.005 & 6450 & A & 1.44e-14\\
4653016974065811328 & NB& 0.261 & 16.719$\pm$0.047 & -1.857$\pm$0.0 & 18.576$\pm$0.047 & 0.013 & 0.71847 & 0.55 & 150 & 0.996 & 6500 & A & 3.24e-15\\
4657937043798067072 & NB& 0.173 & 17.117$\pm$0.208 & -1.63$\pm$0.0 & 18.747$\pm$0.208 & 0.00014 & 0.75115 & 0.6 & 150 & 0.978 & 6650 & A & 1.85e-23\\
4651606468133422208 & NB& 0.48 & 16.299$\pm$0.084 & -2.369$\pm$0.0 & 18.668$\pm$0.084 & 0.013 & 0.71847 & 0.5 & 200 & 1.11 & 6200 & C & 7.96e-26\\
4659781646326951040 & NB& 0.158 & 16.997$\pm$0.03 & -1.603$\pm$0.0 & 18.6$\pm$0.03 & 0.00014 & 0.75115 & 0.6 & 150 & 0.967 & 6700 & A & 1.43e-39\\
\hline
\multicolumn{14}{c}{Silver sample}\\
\hline
4658355751561134848 & NB& 0.328 & 16.683$\pm$0.06 & -2.0$\pm$0.0 & 18.683$\pm$0.06 & 0.00043 & 0.75041 & 0.55 & 200 & 1.055 & 6550 & A & 2.61e-09\\
4657921410118192128 & B& 0.314 & 16.831$\pm$0.102 & -2.275$\pm$0.0 & 19.106$\pm$0.102 & 0.00135 & 0.74806 & 0.75 & 250 & 1.097 & 6550 & A & 1.63e-09\\
4658031365618638592 & NB& 0.427 & 16.588$\pm$0.142 & -2.635$\pm$0.0 & 19.223$\pm$0.142 & 0.00834 & 0.73032 & 0.8 & 300 & 1.16 & 6400 & A & 1.55e-09\\
4651784215395075072 & NB& 0.532 & 16.065$\pm$0.044 & -2.764$\pm$0.0 & 18.829$\pm$0.044 & 0.013 & 0.71847 & 0.65 & 300 & 1.181 & 6250 & D & 5.54e-11\\
4658131073229661184 & NB& 0.418 & 16.489$\pm$0.048 & -2.573$\pm$0.0 & 19.062$\pm$0.048 & 0.013 & 0.71847 & 0.75 & 250 & 1.143 & 6250 & C & 2.47e-11\\
4658763880838210176 & NB& 0.323 & 16.726$\pm$0.08 & -2.031$\pm$0.0 & 18.757$\pm$0.08 & 0.013 & 0.71847 & 0.55 & 150 & 1.046 & 6250 & A & 8.64e-12\\
4757048087051334528 & NB& 0.489 & 16.25$\pm$0.068 & -2.802$\pm$0.0 & 19.052$\pm$0.068 & 0.013 & 0.71847 & 0.75 & 300 & 1.191 & 6200 & C & 3.78e-12\\
4651261839964199552 & B& 0.347 & 16.656$\pm$0.049 & -2.139$\pm$0.0 & 18.795$\pm$0.049 & 0.00834 & 0.73032 & 0.55 & 200 & 1.056 & 6500 & B & 1.3e-12\\
4659206773546904576 & NB& 0.44 & 16.223$\pm$0.06 & -2.663$\pm$0.0 & 18.886$\pm$0.06 & 0.013 & 0.71847 & 0.75 & 300 & 1.156 & 6400 & A & 1.3e-12\\
4657638977349713664 & NB& 0.364 & 16.498$\pm$0.071 & -2.3$\pm$0.0 & 18.798$\pm$0.071 & 0.013 & 0.71847 & 0.65 & 200 & 1.088 & 6300 & C & 1.09e-12\\
4650514172051326208 & B& 0.259 & 16.773$\pm$0.078 & -2.019$\pm$0.0 & 18.792$\pm$0.078 & 0.00424 & 0.74073 & 0.7 & 150 & 1.047 & 6150 & D & 7.87e-13\\
4656279667438493952 & B& 0.259 & 16.816$\pm$0.05 & -1.931$\pm$0.0 & 18.747$\pm$0.05 & 0.013 & 0.71847 & 0.6 & 150 & 1.007 & 6400 & C & 1.89e-13\\
5278221481999140096 & NB& 0.44 & 16.987$\pm$0.263 & -2.502$\pm$0.0 & 19.489$\pm$0.263 & 0.013 & 0.71847 & 0.65 & 250 & 1.134 & 6350 & A & 9.84e-15\\
4657089492111499776 & B& 0.245 & 16.831$\pm$0.07 & -1.904$\pm$0.0 & 18.735$\pm$0.07 & 0.00834 & 0.73032 & 0.6 & 150 & 1.014 & 6400 & C & 9.95e-16\\
4652214777309571328 & B& 0.51 & 16.257$\pm$0.041 & -2.22$\pm$0.0 & 18.477$\pm$0.041 & 0.013 & 0.71847 & 0.5 & 100 & 1.106 & 5300 & C & 3.17e-19\\
4664714880120551680 & B& 0.134 & 17.114$\pm$0.091 & -1.563$\pm$0.0 & 18.677$\pm$0.091 & 0.013 & 0.71847 & 0.6 & 100 & 0.939 & 6300 & C & 7.5e-20\\
4663623924066594944 & NB& 0.266 & 16.78$\pm$0.038 & -1.824$\pm$0.0 & 18.604$\pm$0.038 & 0.00834 & 0.73032 & 0.6 & 100 & 1.014 & 5850 & C & 1.01e-23\\
4660479634388868992 & NB& 0.252 & 16.605$\pm$1.73 & -1.857$\pm$0.0 & 18.462$\pm$1.73 & 0.013 & 0.71847 & 0.55 & 150 & 0.996 & 6500 & A & 1.88e-29\\
\hline
\end{tabular}}
\tablefoot{
        \small $^{\dagger}$ For equivalent [Fe/H] range, see Table~\ref{tab:composition}.}
\label{tab:similar_lc}
\end{table*}

\subsection{BL~Her stars with bumps}

For a comparison of the BL~Her stars with bumps in their $G$ band light curves, we use the complete set of BL~Her models without any constraint on their higher order Fourier amplitude parameters ($R_{k1}$). We have the same constraints on the models with respect to their pulsation periods and amplitudes as in Eq.~\ref{eq:periodconstraint}. The parameter $d$ that is used to quantify the offsets for each modeled-observed pair with respect to their pulsation periods and over their Fourier parameter space remains the same as in Eq.~\ref{eq:d}. However, the different parameter spaces are defined as follows:
\begin{equation}
p \in \{ \log(P), A, S_k, A_c, R_{21}, R_{31}, \phi_{21}, \phi_{31} \}
\end{equation}
for the observed light curves fitted with the optimum order of fit from Baart's condition and
\begin{equation}
p \in \{ \log(P), A, S_k, A_c, R_{21}, R_{31}, R_{41}, R_{51}, R_{61}, R_{71}, \phi_{21}, \phi_{31} \}
\label{eq:N7}
\end{equation}
for the observed light curves fitted with $N=7$. For both cases, the amplitude ($A$) and the skewness ($S_k$) of the light curve have a higher weight of $n_p=10$ while the rest of the parameters have equal weights of $n_p=1$.

Using as many light curve parameters as possible in determining $d$ results in obtaining more precise and accurate modeled-observed pairs. Ideally, we want to include higher order Fourier parameters for all the cases. However, limitations on the observed light curves (for example, not enough epochs or gaps in the epoch photometry) do not always allow us to fit observed light curves with a high order of harmonics and we limit our conditions to include up to $R_{31}$ and $\phi_{31}$ in such cases. For good quality light curves (and especially to capture the bump feature), we fitted the observed light curves with $N=7$, wherever possible and thereby used parameters up to $R_{71}$ (in Eq.~\ref{eq:N7}). Since the bump feature is already well-captured by higher-order Fourier \textit{amplitude} parameters, we have not included the higher-order Fourier \textit{phase} parameters. In Fig.~\ref{fig:phi}, for the particular case of $Gaia$ DR3 4657865541316025600, we observe the need for including higher-order Fourier amplitude parameters up to $R_{71}$ while not necessarily including higher-order Fourier phase parameters up to $\phi_{71}$.

\subsection{Obtaining the best-matched modeled-observed pairs}
\label{sec:best-matched}

Using the above-mentioned conditions, we obtain the score $d$ for each modeled-observed pair. The smaller the score $d$, the smaller the offset for each modeled-observed pair with respect to their pulsation periods and over their Fourier parameter space. Thereafter, for each of the 58 BL~Her stars, we shortlist 10 corresponding models across all four sets of convection parameters that have the smallest $d$ scores. Note that while most stars had at least 10 corresponding models each, a few stars did not, owing to period and amplitude constraints. We investigate the degeneracy in the input stellar parameters of the BL Her models resulting in similar pulsation periods and light curve structure in Section~\ref{sec:degeneracy}; however, for estimating the stellar parameters and the distance to the LMC, we proceed with the \textit{single best} modeled-observed pairs.

In the final stage of the model fitting, the score $d$ has small differences among the 10 best modeled-observed pairs corresponding to a particular observed BL~Her star. For this reason, we choose the final best-fit via the Kolmogorov-Smirnov goodness-of-fit test \citep{kolmogorov1933,smirnov1948}, which tests whether the normalized residuals are from a standard normal distribution \citep{andrae2010}. This in practice means that we calculate the residuals from the observed ($o_i$) and theoretical ($m_i$) light curves, and normalize it with the corresponding observational error ($\sigma_i$):
\begin{equation}
    r_i = \frac{o_i - m_i}{\sigma_i}
\end{equation}

Under the hypothesis that the errors originate from observational errors only, the $r_i$ sample has to follow normal distribution (with 0 mean and standard deviation of 1) for a perfect fit. In practice, due to the fact that the calculation is done on a model grid and secondary light curve features are difficult to be modeled \citep{marconi2017b}, most of the cases do not have a goodness-of-fit above acceptance level. However, the modeled-observed pair that exhibits the largest goodness-of-fit score can be chosen as the best fit result \citep{andrae2010}.

Finally, we verify the best-matched modeled-observed pairs corresponding to the 58 BL~Her stars in the LMC by visual inspection. As mentioned earlier, while the complete set of BL~Her models were used for comparison with the BL~Her stars in the LMC, we choose the lower stellar mass models ($M \leq 0.65\,M_{\odot}$) as a better fit to the observed stars, wherever possible. The light curves of the 30 accepted modeled-observed pairs with superior model fitting are displayed in Fig.~\ref{fig:LC_gold} and labelled \textit{gold sample} for further analysis. Another 18 accepted modeled-observed pairs but with inferior model fitting (especially with respect to their amplitudes or skewness) are displayed in Fig.~\ref{fig:LC_silver} and labelled \textit{silver sample}. For the models that exhibit slightly higher amplitudes as compared to observed light curves, increasing the mixing length parameter may result in smaller pulsation amplitudes \citep[see][among others]{criscienzo2004, paxton2019}. The uncertainties on the convection parameters also affect the theoretical light curve amplitudes \citep{fiorentino2007}. We therefore plan to calibrate the convection parameters for BL~Her models in a future project as was done for RR~Lyrae stars \citep{kovacs2023, kovacs2024}. The light curves of 10 modeled-observed pairs rejected from further analysis are shown in Fig.~\ref{fig:rejected_LC} in the Appendix. An important thing to note here is that this analysis uses the four convection parameter sets as outlined in \citet{paxton2019}; however, these convection parameter values are merely useful starting choices and may be reasonably changed to obtain better fitted models for a particular observed light curve. It is therefore potentially possible to obtain \textsc{mesa-rsp} models with much better fits for the 10 rejected BL~Her stars in Fig.~\ref{fig:rejected_LC} and those will be targets of a future study. 

The estimated stellar parameters from the best-fit models corresponding to the 48 BL~Her stars in the LMC are presented in Table~\ref{tab:similar_lc}. These include the chemical composition ($ZX$), stellar mass ($M$), stellar luminosity ($L$), stellar radius ($R$), effective temperature $T_{\rm eff}$, the convection parameter set used as well as the measure of the final goodness-of-fit score for each modeled-observed pair. An interesting result is that the majority of the best-matched models corresponding to observed BL~Her stars in the LMC are computed using either the convection parameter sets A or C, with very few models computed using convection parameter sets B or D. Note that the convection parameter sets A and C are those without the radiative cooling. The other property that can be important about this parameter selection is that these two sets have the lowest eddy viscosity parameter \citep[refered as $\alpha_m$ in ][]{paxton2018}. This parameter is $0.25$ in set~A and $0.4$ in set~C, while it is $0.5$ in set~B and $0.7$ in set~D. The eddy viscosity parameter controls the amount of kinetic energy converted into turbulent energy \citep{smolec2008}, thereby decreasing the velocity amplitude \citep{kovacs2023}, and also has a significant role in the damping of the pulsation \citep{kovacs2024}. Disentangling the effects of various convective parameters on the BL~Her light curves is beyond the scope of this present study and is planned for the future. We note in passing that lowering the eddy viscosity parameter may also lead to the occurrence of more complex dynamical behavior in non-linear models as well, ranging from period doubling to low-dimensional chaos. This was detected not only in BL~Her but also in W~Vir and RR~Lyrae models \citep{smolec-2012,smolec-2014,smolec-2016,kollath-2011,molnar-2012}.

\begin{figure*}
\centering
\includegraphics[scale = 0.95]{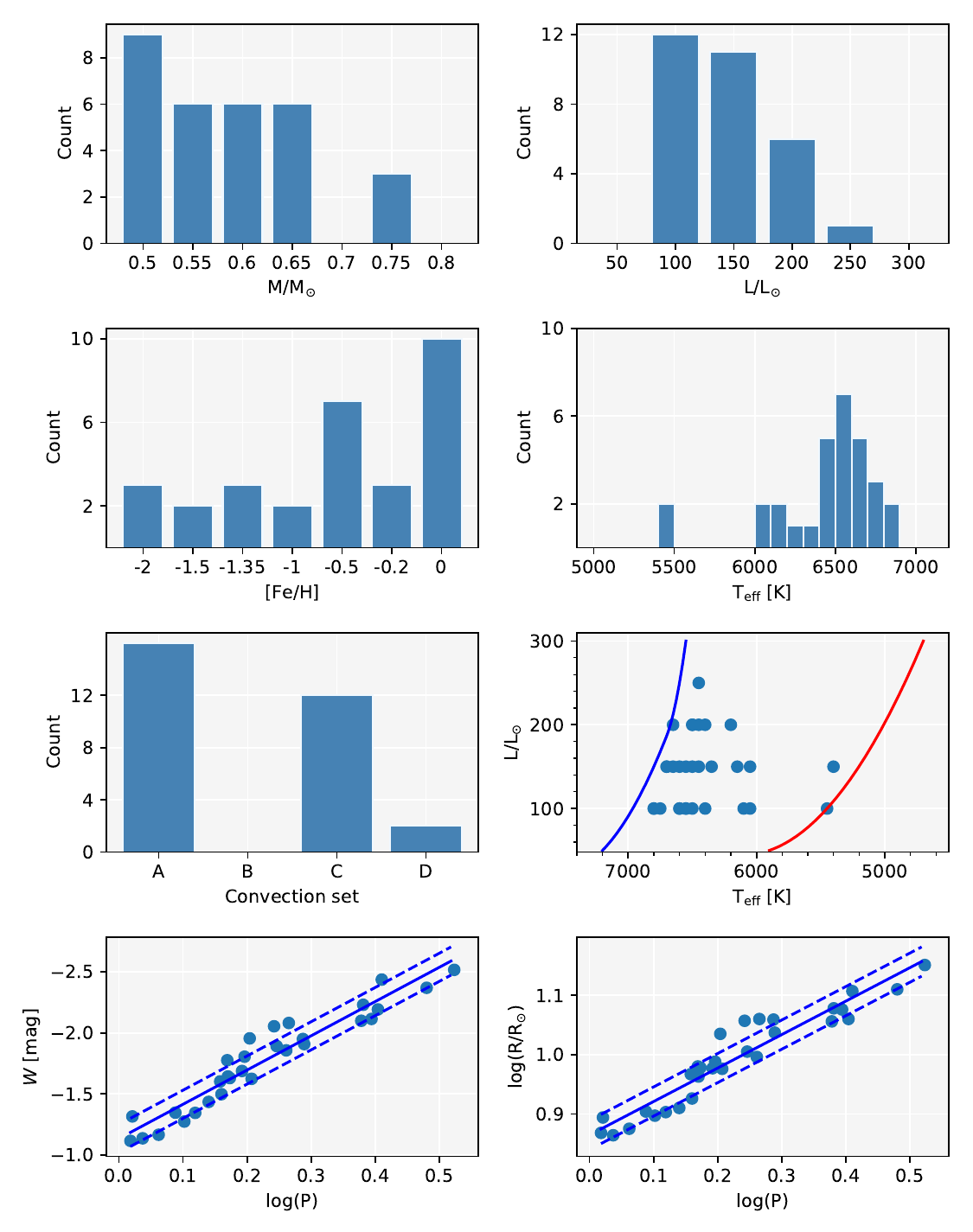}
\caption{A visualisation of the properties of the \textit{gold sample} of 30 BL~Her models that match the best with BL~Her stars in the LMC. The top panels present histograms plotting the distribution of the BL~Her models as a function of stellar mass $M/M_{\odot}$, stellar luminosity $L/L_{\odot}$, metallicity [Fe/H], effective temperature $T_{\rm eff}$ and the convection parameter sets respectively. The location of the subset of BL~Her models is also depicted on the Hertzsprung-Russell diagram; the red and the blue edges of the instability strip are estimated from the linear stability analysis from \citetalias{das2021}. The lowest panels display the period-Wesenheit relation and the period-radius relation of the 30 BL~Her models, respectively. }
\label{fig:table_properties}
\end{figure*}

We further investigate the distribution of the subset of the \textit{gold sample} of 30 BL~Her models that match the best with BL~Her stars in the LMC as a function of metallicity [Fe/H], stellar mass $M/M_{\odot}$, stellar luminosity $L/L_{\odot}$, effective temperature and the convection parameter sets in Fig.~\ref{fig:table_properties}. As noted above, the majority of the models (93.3\%) prefer the convection parameter sets A and C. In \citetalias{das2024}, we found the observed Fourier parameters in the $G$ band from the BL~Her stars in the LMC to overlap well with the theoretical Fourier parameter space from the low-mass models; a detailed light curve comparison between models and observations reveals that 90\% of the best-matched BL~Her models are low-mass ($\leq 0.65 M_{\odot}$) models. Only one model in the subset of best-matched BL~Her models reach stellar luminosity higher than $200 L_{\odot}$. 66.7\% of these BL~Her models belong to the regime of the high-metallicity models ($\rm [Fe/H] \geq -0.5$). Most of the BL~Her models lie near the blue edge of the instability strip, with 63.3\% of the models having effective temperatures between 6400-6700~K. A similar distribution of the subset of the \textit{gold} and \textit{silver} samples of the BL~Her models that match the best with BL~Her stars in the LMC as as a function of [Fe/H], $M/M_{\odot}$, $L/L_{\odot}$, $T_{\rm eff}$ and the convection parameter sets is displayed in Fig.~\ref{fig:table_properties2}.

An alternate classification of pulsating variables with periods between 1 and 3 days was provided by \citet{diethelm1983, diethelm1990} and \citet{sandage1994} and a possible correlation between light curve type and mean metallicity was probed thereafter.  These stars were called above-horizontal-branch (AHB) variables and based on the light curve morphology, they were sub-divided into three classes -- AHB1, AHB2 and AHB3 \citep[for more details, see example light curves provided in][]{diethelm1983}. The 6 BL~Her stars with bumps in our \textit{gold sample} exhibit their bump features on the ascending branch of the light curves, typical of the AHB2 class. From Table~\ref{tab:similar_lc}, we find that 3 of them have $Z=0.00424$ ($[Fe/H]=-0.5$) and another has $Z=0.00834$ ($[Fe/H]=-0.2$). This is in agreement with \citet{diethelm1990} where they found the AHB2 stars to be moderately metal-poor with $-0.7 \leq [Fe/H] \leq -0.1$.  However, our sample also has two bump light curves that have estimated metallicities outside of this range with $Z=0.00043$ ($[Fe/H]=-1.5$) and $Z=0.013$ ($[Fe/H]=0$).

\section{Validating estimated BL~Her models with the OGLE database}
\label{sec:ogle}
\begin{figure*}
\centering
\includegraphics[scale = 1]{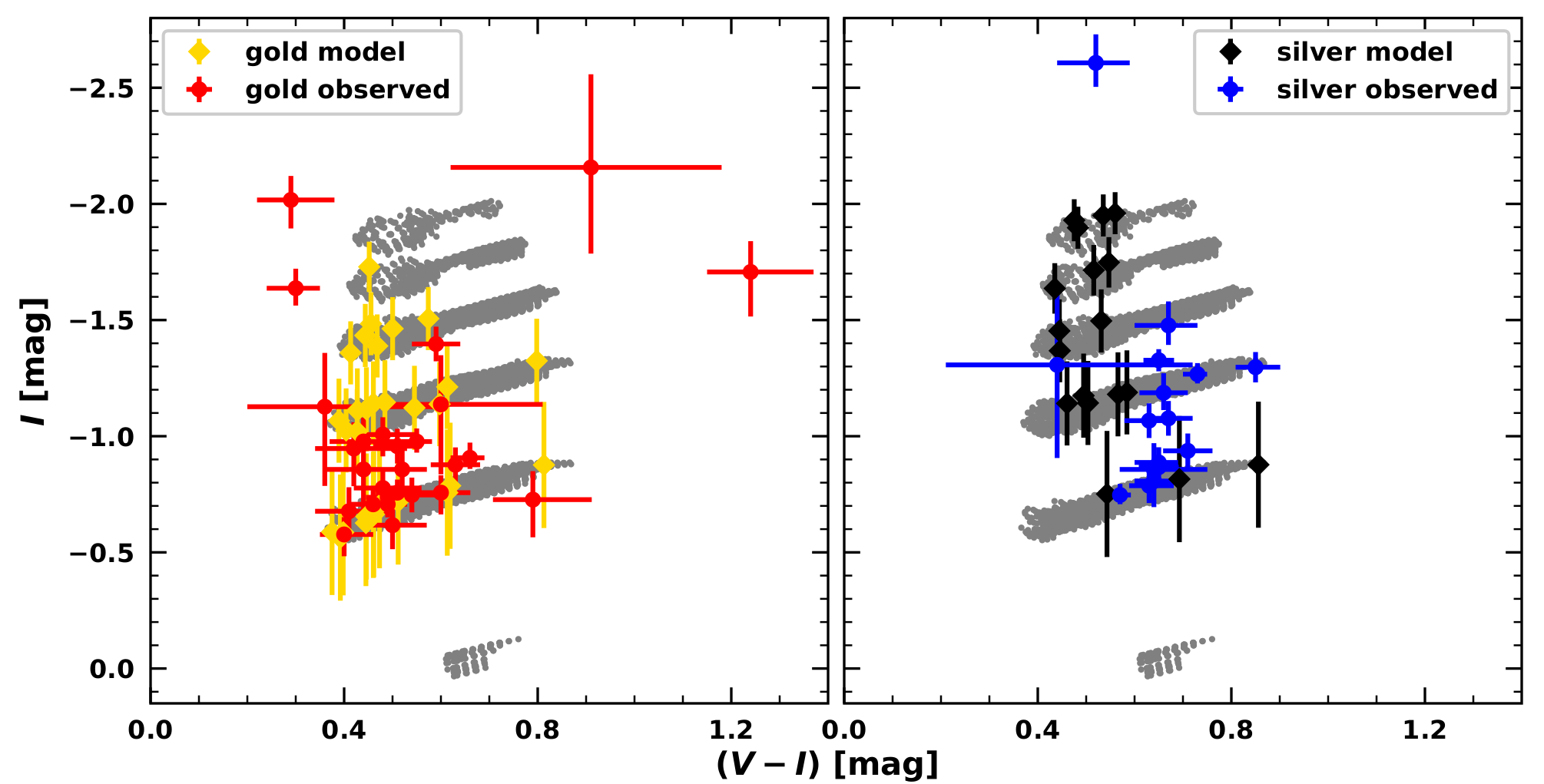}
\caption{The colour-magnitude diagram of the \textit{gold sample} (left) and the \textit{silver sample} (right) of the BL~Her modeled-observed pairs using the OGLE counterparts. The grey points in the background represent all BL~Her models across the four sets of convection parameters.}
\label{fig:CMD}
\end{figure*}

To validate the stellar parameters estimated for the observed BL~Her stars in the LMC (see Table~\ref{tab:similar_lc}), we study the colour-magnitude diagram (CMD) of the modeled-observed pairs but using the OGLE observations of the observed BL~Her stars. This is presented in Fig.~\ref{fig:CMD}. We make use of the optical ($VI$) light curves of T2Cs (and classical Cepheids) in the LMC from the OGLE-IV catalog to obtain the observed CMD \citep{soszynski2015, soszynski2018}. 
To account for interstellar extinction, we use the $E(V-I)$ color excess values from the reddening map of \citet{skowron2021}, which is designed specifically for the Magellanic Clouds. We then transform those to $E(B-V)$  color excess values using $E(V-I) = 1.38 \times E(B-V)$ \citep{tammann2003}. We finally obtain the extinction coefficients $A_V,A_I$ using the following conversion factors from \citet{schlegel1998}:

\begin{equation}
\begin{aligned}
A_V &= 3.32 \times E(B-V), \\
A_I &= 1.94 \times E(B-V).
\end{aligned}
\end{equation}

The extinction-corrected $I$ band apparent magnitudes are then converted to their respective absolute magnitudes using a distance modulus of $\mu_{\rm LMC} = 18.477 \pm 0.004$ (statistical) $\pm 0.026$ (systematic) from \citet{pietrzynski2019}, which is the present best-estimate of the geometrical distance to the LMC obtained by using 20 eclipsing binary systems and is precise to 1 per cent. The uncertainties on the observed CMD take into account errors from the reddening map, the errors on the mean $VI$ magnitudes, and the errors on the estimated LMC distance modulus. 

The theoretical CMD is obtained by using the absolute $VI$ magnitudes of the best-fit models \citepalias[see][]{das2021}, and the associated errors are estimated using Gaussian error propagation, with a luminosity step size of $\pm 25\,L_{\odot}$ of the computed grid of BL~Her models. A comparison of the theoretical and observed $I$ band light curves of the \textit{gold sample} is displayed in Fig.~\ref{fig:LC_gold_ogle} in the Appendix.

From Fig.~\ref{fig:CMD}, we find that the \textit{gold sample} of 30 BL~Her modeled-observed pairs in the $(V-I)$~vs~$I$ CMD plane match reasonably well, especially for the bluer stars with $(V-I) < 0.62$. While BL~Her models do exist for  $(V-I) > 0.62$, we do not find good matches with the observed BL~Her stars in this range. This seems like a current limitation of our model-fitting technique, where parameters for bluer stars are estimated better than that of red stars. Note that the choice of convection parameters plays a huge role in the stellar pulsation models (and subsequently their light curves), especially towards the red edge of the instability strip \citep{deupree1977a, stellingwerf1982a, stellingwerf1982b}. The comparison of the theoretical and observed CMD of the \textit{silver sample} of 18 BL~Her modeled-observed pairs show that the BL~Her models are 
offset from the observed BL~Her stars. We therefore proceed with the \textit{gold sample} of 30 BL~Her modeled-observed pairs only in the subsequent analysis for distance estimation to the LMC.

\section{Estimating the distance to the LMC}
\label{sec:distance}

\begin{figure*}
\centering
\includegraphics[scale = 1]{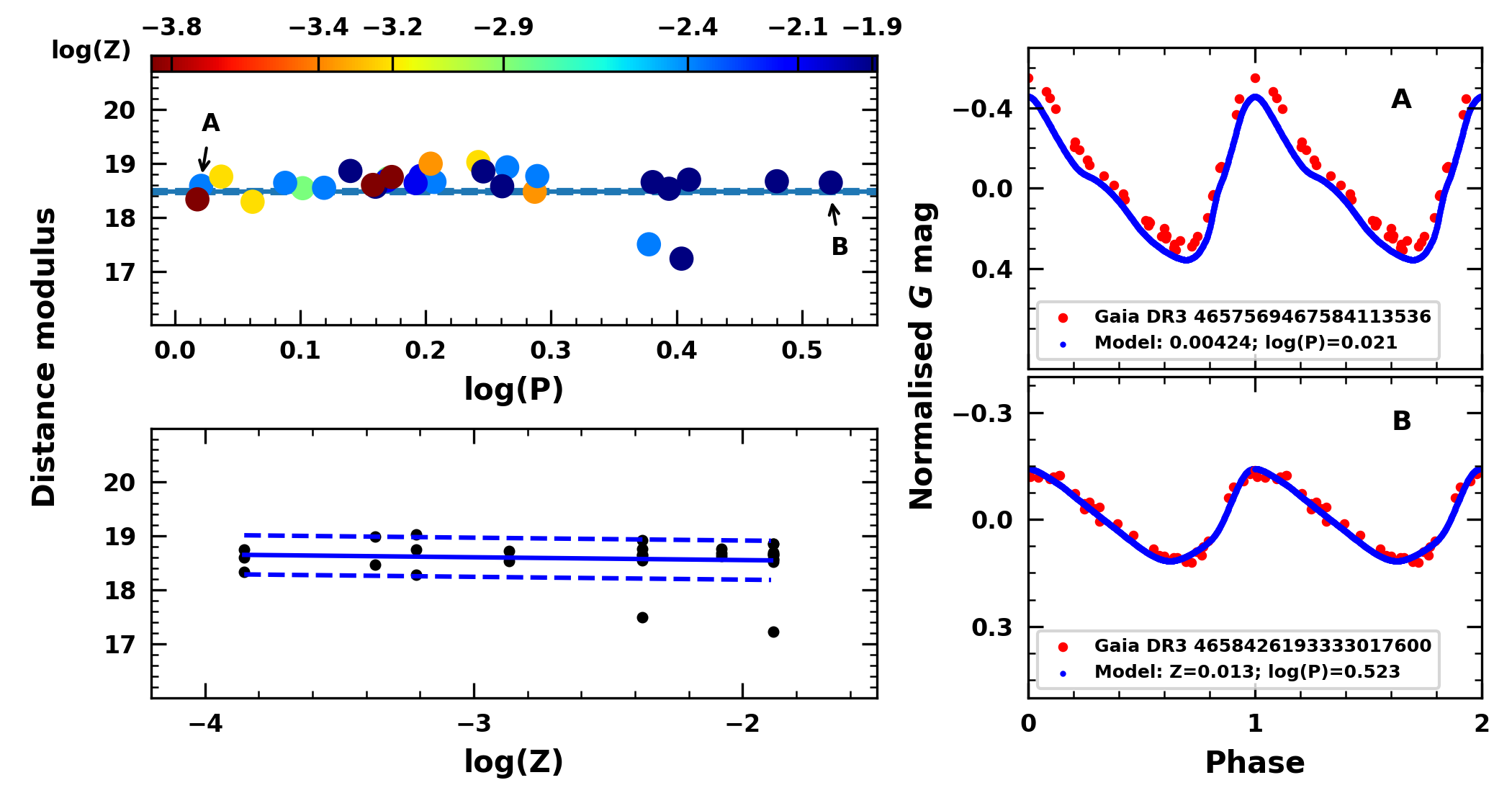}
\caption{The left-hand panels display the variation of distance modulus as a function of period (top) and metal abundance (bottom). Solid lines represent $\mu_{\rm LMC} = 18.477 \pm 0.026$ \citep{pietrzynski2019} in the top panel and best-fitting linear regression in the bottom panel while the dashed lines represent $1\sigma$ error in each case. The right-hand panel shows theoretical and observed (normalized with respect to mean magnitude) $G$ band light curves of two BL~Her stars that are consistent on the Fourier plane.}
\label{fig:distmod}
\end{figure*}

As a direct application of our comparative analysis between theoretical and observed light curves of BL~Her stars, we use the absolute Wesenheit magnitudes from the models and the apparent Wesenheit magnitudes from the observed BL~Her stars in the LMC to estimate the distance modulus to the LMC.

As mentioned before, the comparison of the theoretical light curves with those from the observed BL~Her stars in the LMC was carried out using the $G$ band light curves only, since the empirical light curves contain better photometric data with a good phase coverage as compared to those in the $G_{BP}$ and $G_{RP}$ passbands. We have not corrected for extinction in this present study. The light curve structure itself is not affected by extinction correction, or lack thereof. However, comparing the observed apparent magnitudes without extinction correction with the absolute magnitudes from the models using the light curves in the $G$ band only would affect our estimation of the distance modulus to the LMC. In light of this, we make use of the modeled-observed pairs that exhibit best matches with respect to their pulsation periods and $G$ band light curve structure but obtain their corresponding Wesenheit magnitudes, as defined by \citet{ripepi2019} for $Gaia$ DR3:
\begin{equation}
 W(G, G_{BP} - G_{RP})=G - 1.90(G_{BP}-G_{RP}), 
 \label{eq:wesenheit}
\end{equation}
where $G$, $G_{BP}$, and $G_{RP}$ are the mean magnitudes in the respective $Gaia$ passbands. The Wesenheit magnitudes for the observed BL~Her stars are calculated using the mean $GG_{BP}G_{RP}$ magnitudes as provided in the $Gaia$~DR3 database\footnote{\url{https://gea.esac.esa.int/archive/}}.

Wesenheit magnitudes offer the advantage of being minimally affected by the uncertainties arising from reddening corrections \citep{madore1982}. However, they are quite sensitive to the extinction law assumed. The extinction law of the LMC has been quite extensively studied \citep[][among others]{gordon2003, gao2013, wang2023}, thereby justifying our use of Wesenheit magnitudes as reddening-free magnitudes in this work. The Wesenheit magnitudes obtained from the theoretical and the observed light curves for the best-matched modeled-observed pairs are included in Table~\ref{tab:similar_lc}; these are thereby used to estimate the individual distance moduli $\mu$ to the LMC. The distribution of the individual distance moduli as a function of the pulsation period and the metal abundance is displayed in the upper left-hand panel of Fig.~\ref{fig:distmod}. The variation of the distance moduli with metal abundance (slope = $-0.052 \pm 0.103$ with a scatter of $\sigma =0.363$) is presented in the lower left-hand panel of Fig.~\ref{fig:distmod}; however, given the small sample, we cannot draw any conclusive evidence of a possible correlation between the two.
 
Finally, we obtain an average distance modulus to the LMC of $\mu_{\rm LMC} = 18.582 \pm 0.067$ using the \textit{gold sample} of 30 best-matched modeled-observed pairs, as summarised in Table~\ref{tab:similar_lc}. As mentioned before, \citet{pietrzynski2019} provides the most accurate and precise measurement of the geometrical distance to the LMC to date of $\mu_{\rm LMC} = 18.477 \pm 0.004$ (statistical) $\pm 0.026$ (systematic). \citet{wielgorski2022} used the $PL$ relations of Milky Way T2Cs to obtain a distance modulus of $\mu_{\rm LMC} = 18.540 \pm 0.026$ (statistical) $\pm 0.034$ (systematic). Our results are therefore in reasonably good agreement with published results and lie within the bounds of the above-quoted literature values.

The period-Wesenheit relation for the subset of the best-matched \textit{gold sample} of 30 BL~Her models is displayed in the lowest panel of Fig.~\ref{fig:table_properties} and exhibits a slope of $-2.805 \pm 0.164$ with a scatter of $\sigma = 0.115$. For comparison, the empirical period-Wesenheit relation for the 58 BL~Her stars in the LMC has a slope of $-2.398\pm0.146$ with a scatter of $\sigma = 0.144$ (see Table~5 of \citetalias{das2024}). The $T$ statistic\footnote{We defined a $T$ statistic for the comparison of two linear regression slopes, $\hat{W}$ with sample sizes, $n$ and $m$, respectively, as follows:
\begin{equation}\nonumber
T=\frac{\hat{W}_n-\hat{W}_m}{\sqrt{\mathrm{Var}(\hat{W}_n)+\mathrm{Var}(\hat{W}_m)}},
\label{eq:ttest}
\end{equation}
where $\mathrm{Var}(\hat{W})$ is the variance of the slope. The null hypothesis of equivalent slopes is rejected if $T>t_{\alpha/2,\nu}$ (or the probability of the observed value of the $T$ statistic) is $p<0.05,$ where $t_{\alpha/2,\nu}$ is the critical value under the two-tailed $t$-distribution with 95\% confidence limit ($\alpha$=0.05) and degrees of freedom, $\nu=n+m-4$.} and its associated probability $p(t)$ of acceptance of the null hypothesis (equal slopes) obtained for the comparison of the two linear regression slopes under the two-tailed $t$-distribution are 1.854 and 0.067 respectively, thereby suggesting statistically similar $PW$ relations.  

The period-radius relation for the subset of the best-matched \textit{gold sample} of 30 BL~Her models is also presented in the lowest panel of Fig.~\ref{fig:table_properties} and exhibits a slope of $0.565 \pm 0.035$ with a scatter of $\sigma = 0.025$. This is in excellent agreement with the empirical $PR$ slope of $0.564 \pm 0.049$ with a scatter of $\sigma = 0.047$ from the BL~Her stars in the LMC \citep{groenewegen2017b}. The (|$T$|, $p(t)$) values for the comparison of the two linear regression slopes are (0.017, 0.987).

\section{One star--many models}
\label{sec:degeneracy}

\begin{figure*}
\centering
\includegraphics[width = 1\textwidth]{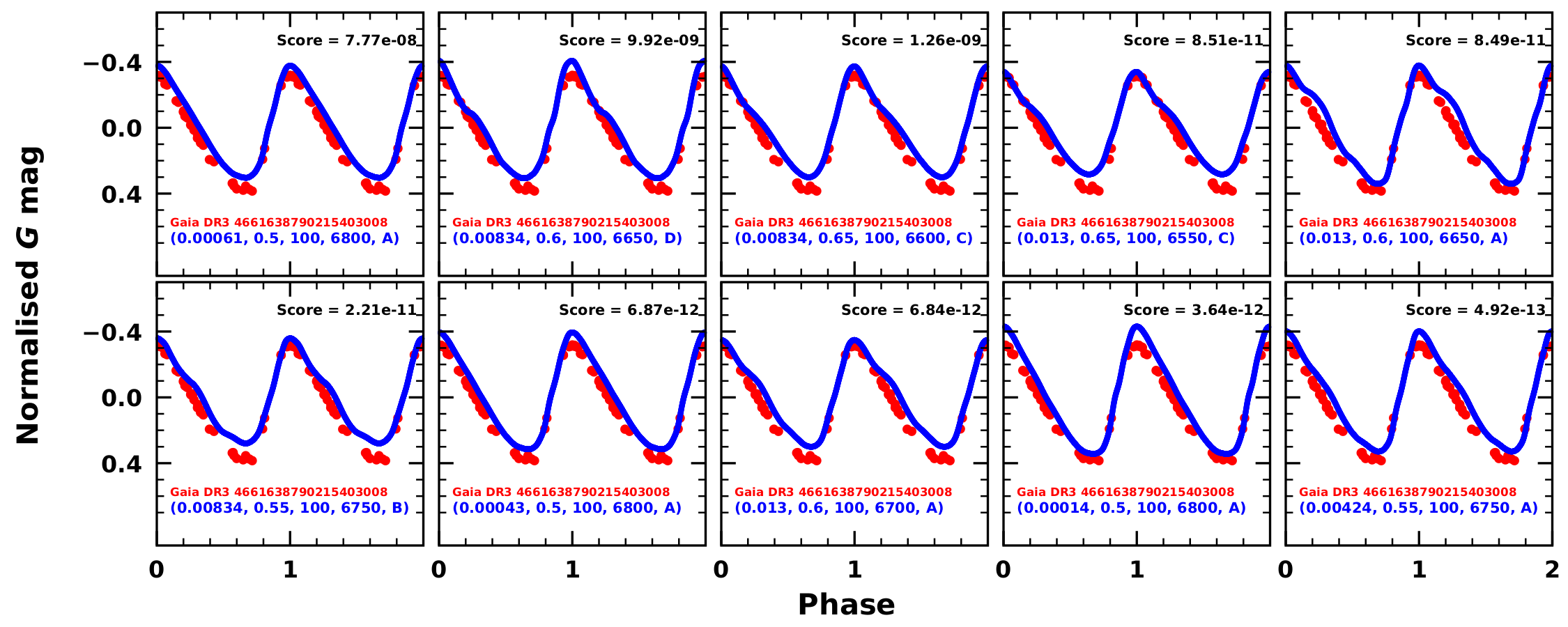}
\caption{An example of the 10 best-matched models corresponding to one particular observed BL~Her star Gaia~DR3~4661638790215403008. The observed light curve is shown in red while the theoretical light curves are displayed in blue. The input stellar parameters of the corresponding models are included in the format ($Z, M/M_{\odot}, L/L_{\odot}, T_{\rm eff}$, convection set) in each sub-plot.}
\label{fig:3008}
\end{figure*}

\begin{figure*}
\centering
\includegraphics[scale = 1]{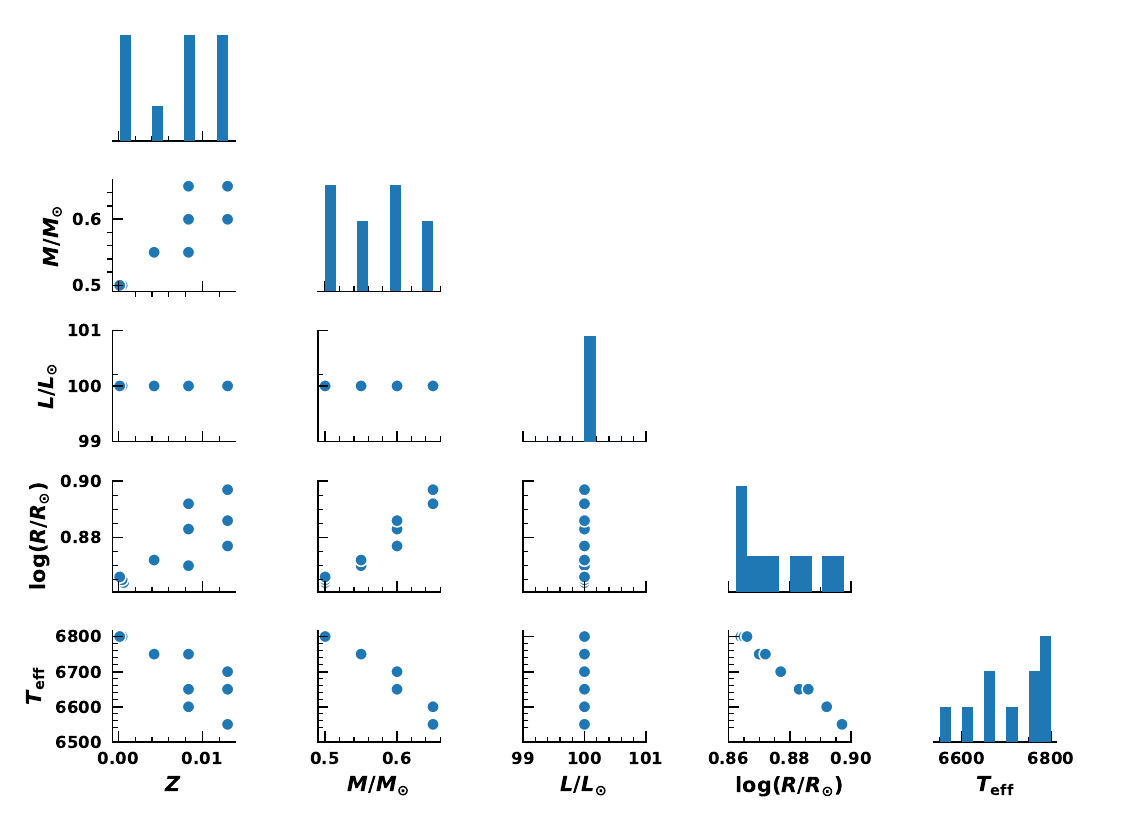}
\caption{A corner plot exhibiting the correlations of the different stellar parameters (chemical composition $Z$, stellar mass $M/M_{\odot}$, stellar luminosity $L/L_{\odot}$, stellar radius $\log(R/R_{\odot})$ and effective temperature $T_{\rm eff}$) using the 10 best-matched models corresponding to one particular observed BL~Her star Gaia~DR3~4661638790215403008.}
\label{fig:onestarmanymodels}
\end{figure*}

\begin{figure*}
\centering
\includegraphics[width = 1\textwidth]{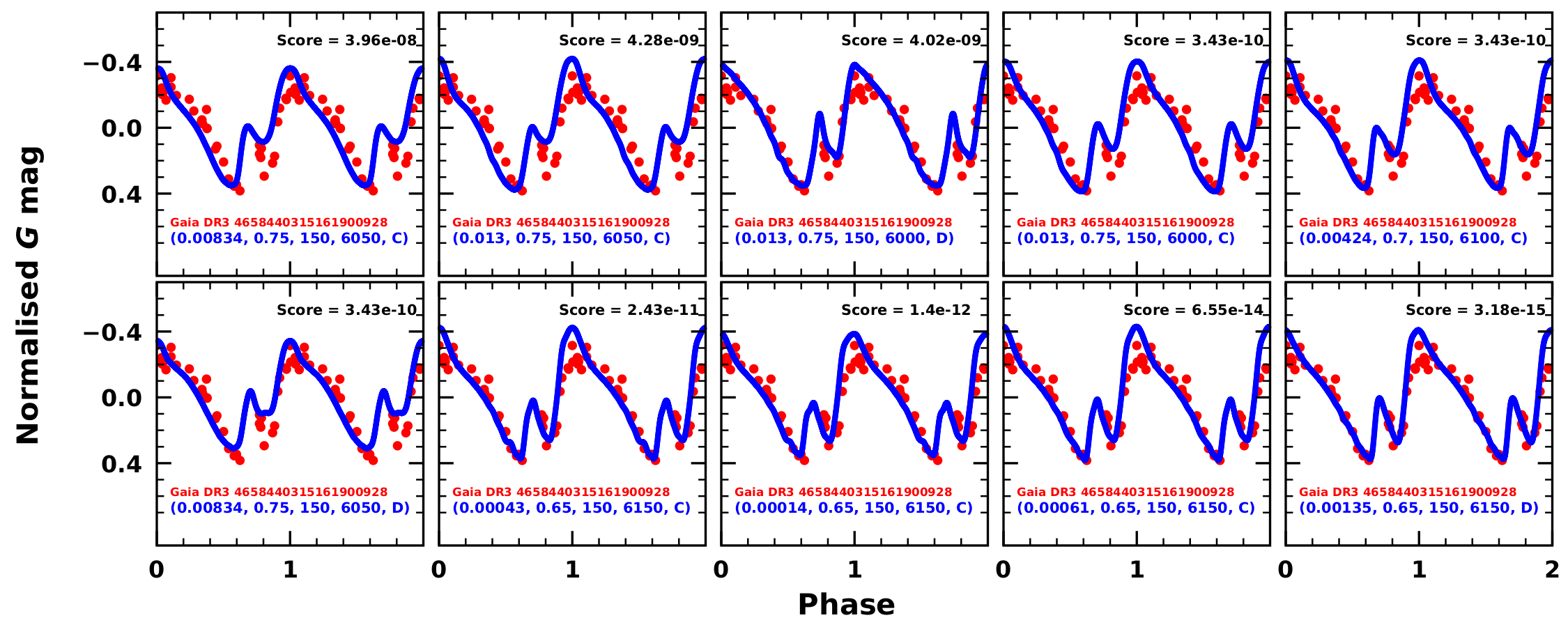}
\caption{Same as Fig.~\ref{fig:3008} but for the observed BL~Her star with the bump feature Gaia~DR3~4658440315161900928.}
\label{fig:0928}
\end{figure*}

\begin{figure*}[h!]
\centering
\includegraphics[scale = 1]{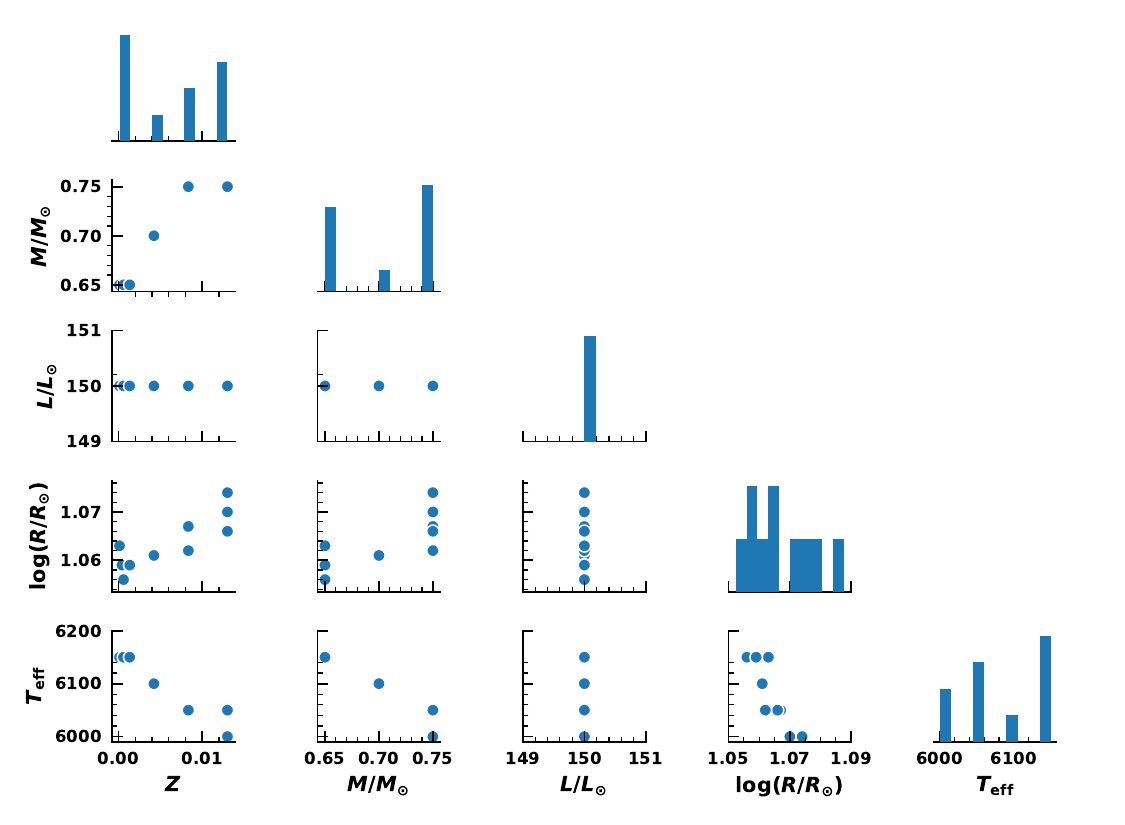}
\caption{Same as Fig.~\ref{fig:onestarmanymodels} but for the observed BL~Her star Gaia~DR3~4658440315161900928. }
\label{fig:onestarmanymodels1}
\end{figure*}

So far, we have discussed the results from the best-matched model corresponding to a particular observed star. However, as mentioned in Section~\ref{sec:best-matched}, we shortlisted 10 models each that match the closest with an observed BL~Her star with respect to their pulsation periods and their Fourier parameter space. In this section, we study the degeneracy in the input stellar parameters of the BL~Her models that result in similar pulsation periods and light curve structure. An example of the 10 best-matched models corresponding to one particular observed BL~Her star without bump, Gaia~DR3~4661638790215403008, is displayed in Fig.~\ref{fig:3008}, along with the input stellar parameters of the corresponding models in each sub-plot. We also study the correlations of the different stellar parameters $-$ chemical composition $Z$, stellar mass $M/M_{\odot}$, stellar luminosity $L/L_{\odot}$, stellar radius $\log(R/R_{\odot})$ and effective temperature $T_{\rm eff}$ $-$ of the 10 best-matched models corresponding to the star in Fig.~\ref{fig:onestarmanymodels}. While the stellar mass is restricted to the range $M/M_{\odot} \in [0.5,0.65]$, with a fixed stellar luminosity $L/L_{\odot}=100$, the effective temperature is quite broad in the range $T_{\rm eff} \in [6550,6800]$ (K). There is also a wide range of chemical compositions with $Z \in \{ 0.00014, 0.00043, 0.00061, 0.00424, 0.00834, 0.013 \}$. However, the majority of the models (80\%) still prefer the convection parameter sets A and C. 

An example of the 10 best-matched models corresponding to the observed BL~Her star with bump, Gaia~DR3~4658440315161900928, and the correlations of the different stellar parameters are displayed in Fig.~\ref{fig:0928} and Fig.~\ref{fig:onestarmanymodels1}, respectively. The stellar mass is restricted to the range $M/M_{\odot} \in [0.65,0.75]$ with a fixed stellar luminosity $L/L_{\odot}=150$ and the effective temperature in the range of $T_{\rm eff} \in [6000,6150]$ (K). The chemical composition cannot be constrained, as the 10 best-matched models span the entire range of the input parameter used in the grid with $Z \in \{ 0.00014, 0.00043, 0.00061, 0.00135, 0.00424, 0.00834, 0.013 \}$. Once again, the majority of the models (70\%) still prefer the convection parameter set C.

This analysis was carried out to show that while the stellar estimates provided in Table~\ref{tab:similar_lc}, using our robust light-curve fitting approach, are reasonably good (especially for the \textit{gold sample} BL~Her modeled-observed pairs), one has to be careful because degeneracies exist in the models that may result in similar pulsation periods and light curve structure. A way to constrain the modeled-observed pairs further would be to simultaneously fit the light curves and the radial velocity curves for stars and models that have similar pulsation periods. The two examples above also suggest that while the metallicity range may be difficult to constrain, one could re-compute new models by adopting the weighted averaged values of $M, L, T_{\rm eff}$ (with the weights scaled according to their goodness-of-fit scores) of the 10 best modeled-observed pairs, keeping the $ZX$ and convection parameter sets the same. Thereafter, a second iteration of the comparison of the observed and the new modeled light curves may help constrain the input stellar parameters (and especially, the metallicity) of the models better. However, we note that this study is the most robust light-curve fitting that can be done at present for BL~Her stars using the latest available observations from $Gaia$ DR3 and the most recent stellar pulsation models. We also plan to constrain the convection parameters for BL~Her models in a future project, similar to what has been carried out for RRab \citep{kovacs2023} and RRc stars \citep{kovacs2024}.

\section{Summary and Conclusion}
\label{sec:results}

We carried out an extensive comparison of the light curve structure of BL~Her stars in the $G$ band using observations from $Gaia$ DR3 \citep{prusti2016, vallenari2023} and the most recent stellar pulsation models computed for a wide range of input parameters $-$ metallicity ($-2.0\; \mathrm{dex} \leq \mathrm{[Fe/H]} \leq 0.0\; \mathrm{dex}$), stellar mass (0.5M$_{\odot}$-0.8M$_{\odot}$), stellar luminosity (50L$_{\odot}$-300L$_{\odot}$) and effective temperature (full extent of the instability strip; in steps of 50K) typical for BL~Her stars using \textsc{mesa-rsp} \citepalias{das2021, das2024}. The important results from this analysis are summarized below:

\begin{enumerate}

\item We compared the $G$ band theoretical light curves of the complete set of BL~Her models computed using the different convection parameter sets with the observed light curves of the BL~Her stars in the LMC that have similar pulsation periods. We used the mathematically robust goodness-of-fit measure over the Fourier parameter space to obtain best-fit models for 48 stars in the LMC and provide the stellar parameter estimates in Table~\ref{tab:similar_lc} of this paper.

\item We find a relatively flat distribution of stellar masses between 0.5--0.65\,$M_\odot$. Although 90\% of models fall into this range, we found three stars where the fits clearly preferred a higher mass of $0.75\,M_\odot$. 

\item An interesting result is that while there was no obvious preference of the choice of convection parameter sets in \citetalias{das2024} when we compared the theoretical and empirical period-Wesenheit relations of BL~Her stars at mean light, an extensive light curve comparison analysis shows that the majority of the best-matched models are computed using the convection parameter sets A and C. Note that the convection parameter sets A and C are computed not only without radiative cooling, but also have low eddy viscosities. The result that the majority of the best-matched models are computed using sets A and C could be an interplay of the different convective parameters, and does not allow us to conclude that radiative cooling is inefficient in these models.

\item The period-Wesenheit relation for the subset of the best-matched \textit{gold sample} of 30 BL~Her models exhibits a slope of $-2.805 \pm 0.164$, which is in good agreement with the empirical $PW$ slope of $-2.398\pm0.146$ from the BL~Her stars in the LMC \citepalias[using data from $Gaia$ DR3 in][]{das2024}. We also found the period-radius relation for the \textit{gold sample} of 30 BL~Her models to exhibit a slope of $0.565 \pm 0.035$ in excellent agreement with the empirical $PR$ slope of $0.564 \pm 0.049$ from the BL~Her stars in the LMC \citep{groenewegen2017b}.

\item We used the Wesenheit magnitudes of the subset of 30 best-matched modeled-observed pairs to estimate a distance modulus to the LMC of $\mu_{\rm LMC} = 18.582 \pm 0.067$ which lies within the bounds of previous literature values from \citet{pietrzynski2019} and \citet{wielgorski2022}.

\item We also analysed the degeneracy in the stellar parameters of the BL~Her models that result in similar pulsation periods and light curve structure and noted that an even more robust method of fitting would be to simultaneously fit multi-band light curves and radial velocity curves for stars and models that have similar pulsation periods. However, we note that this study is the most robust light-curve fitting that can be done at present using the most recent theoretical and observed data.

\item Despite the large number of BL~Her models, we did not find best-fit models for 10 BL~Her stars in the LMC. A possible explanation could be the \textit{curse of dimensionality} \citep{bellman1959} which suggests that the number of configurations (in this work: pulsation period, light curve amplitude and Fourier parameters) covered by an observation decreases as the dimensionality or parameter space of the models increases. Another reason is that this analysis uses the four convection parameter sets as outlined in \citet{paxton2019}; however, these convection parameter values are merely useful starting choices. Our grid of models suggest that the choice of convection parameters clearly play an important role in the theoretical light curve structure, especially on the Fourier amplitude parameters and therefore, may be reasonably changed to obtain better fitted models for a particular observed light curve.

\end{enumerate}

Our analysis successfully demonstrates that in the era of percent-level precision in the calibration of $PL$ relations of classical pulsators, it is extremely important to not just compare the observations and the stellar pulsation models at mean light but also compare their light curves over a complete cycle of pulsation for better constrains on the stellar pulsation theory. The extensive light curve comparison analysis shows a clear preference for radial stellar pulsation models computed without radiative cooling for BL~Her stars using \textsc{mesa-rsp}. However, as mentioned earlier, this does not allow us to conclude that radiative cooling is inefficient in these models and could be the result of an interplay between the various convective parameters. It is therefore important to calibrate the convection parameters as was done for RRab \citep{kovacs2023} and RRc stars \citep{kovacs2024} and thereby we plan to constrain the convection parameters for BL~Her models in a future project. We would like to highlight here that adopting static model atmospheres while the pulsating atmosphere is clearly dynamic may also affect our extensive light curve comparison analysis between observed and theoretical light curves.

\begin{acknowledgements}
The authors thank the referee for useful comments and suggestions that improved the quality of the manuscript. This research was supported by the KKP-137523 `SeismoLab' \'Elvonal grant and by the NKFIH excellence grant TKP2021-NKTA-64 of the Hungarian Research, Development and Innovation Office (NKFIH). M.J.\ gratefully acknowledges funding from MATISSE: \textit{Measuring Ages Through Isochrones, Seismology, and Stellar Evolution}, awarded through the European Commission's Widening Fellowship. This project has received funding from the European Union's Horizon 2020 research and innovation program. RS is supported by the Polish National Science Centre, SONATA BIS grant, 2018/30/E/ST9/00598. This research was supported by the International Space Science Institute (ISSI) in Bern/Beijing through ISSI/ISSI-BJ International Team project ID \#24-603 - “EXPANDING Universe” (EXploiting Precision AstroNomical Distance INdicators in the Gaia Universe). The authors acknowledge the use of High Performance Computing facility Pegasus at IUCAA, Pune as well as at the HUN-REN (formerly ELKH) Cloud and the following software used in this project: \textsc{mesa}~r11701 \citep{paxton2011, paxton2013, paxton2015, paxton2018, paxton2019, jermyn2023}. This research has made use of NASA’s Astrophysics Data System.
\end{acknowledgements}

\bibliographystyle{aa} 

\begin{thebibliography}{101}
\expandafter\ifx\csname natexlab\endcsname\relax\def\natexlab#1{#1}\fi

\bibitem[{{Andrae} {et~al.}(2010){Andrae}, {Schulze-Hartung}, \&
  {Melchior}}]{andrae2010}
{Andrae}, R., {Schulze-Hartung}, T., \& {Melchior}, P. 2010, arXiv e-prints,
  arXiv:1012.3754

\bibitem[{{Asplund} {et~al.}(2009){Asplund}, {Grevesse}, {Sauval}, \&
  {Scott}}]{asplund2009}
{Asplund}, M., {Grevesse}, N., {Sauval}, A.~J., \& {Scott}, P. 2009, ARA\&A,
  47, 481

\bibitem[{Baart(1982)}]{baart1982}
Baart, M.~L. 1982, IMA Journal of Numerical Analysis, 2, 241

\bibitem[{{Beaton} {et~al.}(2016){Beaton}, {Freedman}, {Madore}, {Bono},
  {Carlson}, {Clementini}, {Durbin}, {Garofalo}, {Hatt}, {Jang}, {Kollmeier},
  {Lee}, {Monson}, {Rich}, {Scowcroft}, {Seibert}, {Sturch}, \&
  {Yang}}]{beaton2016}
{Beaton}, R.~L., {Freedman}, W.~L., {Madore}, B.~F., {et~al.} 2016, ApJ, 832,
  210

\bibitem[{{Bellinger} {et~al.}(2020){Bellinger}, {Kanbur}, {Bhardwaj}, \&
  {Marconi}}]{bellinger2020}
{Bellinger}, E.~P., {Kanbur}, S.~M., {Bhardwaj}, A., \& {Marconi}, M. 2020,
  \mnras, 491, 4752

\bibitem[{Bellman \& Kalaba(1959)}]{bellman1959}
Bellman, R. \& Kalaba, R. 1959, IRE Transactions on Automatic Control, 4, 1

\bibitem[{{Bhardwaj} {et~al.}(2017{\natexlab{a}}){Bhardwaj}, {Kanbur},
  {Marconi}, {Rejkuba}, {Singh}, \& {Ngeow}}]{bhardwaj2017a}
{Bhardwaj}, A., {Kanbur}, S.~M., {Marconi}, M., {et~al.} 2017{\natexlab{a}},
  MNRAS, 466, 2805

\bibitem[{{Bhardwaj} {et~al.}(2022){Bhardwaj}, {Kanbur}, {Rejkuba}, {Marconi},
  {Catelan}, {Ripepi}, \& {Singh}}]{bhardwaj2022}
{Bhardwaj}, A., {Kanbur}, S.~M., {Rejkuba}, M., {et~al.} 2022, \aap, 668, A59

\bibitem[{{Bhardwaj} {et~al.}(2017{\natexlab{b}}){Bhardwaj}, {Rejkuba},
  {Minniti}, {Surot}, {Valenti}, {Zoccali}, {Gonzalez}, {Romaniello}, {Kanbur},
  \& {Singh}}]{bhardwaj2017c}
{Bhardwaj}, A., {Rejkuba}, M., {Minniti}, D., {et~al.} 2017{\natexlab{b}},
  A\&A, 605, A100

\bibitem[{{Bono} {et~al.}(2020){Bono}, {Braga}, {Fiorentino}, {Salaris},
  {Pietrinferni}, {Castellani}, {Di Criscienzo}, {Fabrizio},
  {Mart{\'\i}nez-V{\'a}zquez}, \& {Monelli}}]{bono2020}
{Bono}, G., {Braga}, V.~F., {Fiorentino}, G., {et~al.} 2020, A\&A, 644, A96

\bibitem[{{Bono} {et~al.}(1997){Bono}, {Caputo}, \& {Santolamazza}}]{bono1997a}
{Bono}, G., {Caputo}, F., \& {Santolamazza}, P. 1997, A\&A, 317, 171

\bibitem[{{Bono} {et~al.}(2000){Bono}, {Castellani}, \& {Marconi}}]{bono2000d}
{Bono}, G., {Castellani}, V., \& {Marconi}, M. 2000, \apjl, 532, L129

\bibitem[{{Bono} {et~al.}(2002){Bono}, {Castellani}, \& {Marconi}}]{bono2002}
{Bono}, G., {Castellani}, V., \& {Marconi}, M. 2002, ApJ, 565, L83

\bibitem[{{Braga} {et~al.}(2020){Braga}, {Bono}, {Fiorentino}, {Stetson},
  {Dall'Ora}, {Salaris}, {da Silva}, {Fabrizio}, {Marinoni}, {Marrese},
  {Mateo}, {Matsunaga}, {Monelli}, \& {Wallerstein}}]{braga2020}
{Braga}, V.~F., {Bono}, G., {Fiorentino}, G., {et~al.} 2020, A\&A, 644, A95

\bibitem[{{Buchler} \& {Moskalik}(1992)}]{buchler1992}
{Buchler}, J.~R. \& {Moskalik}, P. 1992, ApJ, 391, 736

\bibitem[{{Buchler} {et~al.}(1990){Buchler}, {Moskalik}, \&
  {Kovacs}}]{buchler1990}
{Buchler}, J.~R., {Moskalik}, P., \& {Kovacs}, G. 1990, \apj, 351, 617

\bibitem[{{Caputo}(1998)}]{caputo1998}
{Caputo}, F. 1998, \aapr, 9, 33

\bibitem[{{Das} {et~al.}(2018){Das}, {Bhardwaj}, {Kanbur}, {Singh}, \&
  {Marconi}}]{das2018}
{Das}, S., {Bhardwaj}, A., {Kanbur}, S.~M., {Singh}, H.~P., \& {Marconi}, M.
  2018, MNRAS, 481, 2000

\bibitem[{{Das} {et~al.}(2020){Das}, {Kanbur}, {Bellinger}, {Bhardwaj},
  {Singh}, {Meerdink}, {Proietti}, {Chalmers}, \& {Jordan}}]{das2020}
{Das}, S., {Kanbur}, S.~M., {Bellinger}, E.~P., {et~al.} 2020, MNRAS, 493, 29

\bibitem[{{Das} {et~al.}(2021){Das}, {Kanbur}, {Smolec}, {Bhardwaj}, {Singh},
  \& {Rejkuba}}]{das2021}
{Das}, S., {Kanbur}, S.~M., {Smolec}, R., {et~al.} 2021, \mnras, 501, 875

\bibitem[{{Das} {et~al.}(2024){Das}, {Moln{\'a}r}, {Kanbur}, {Joyce},
  {Bhardwaj}, {Singh}, {Marconi}, {Ripepi}, \& {Smolec}}]{das2024}
{Das}, S., {Moln{\'a}r}, L., {Kanbur}, S.~M., {et~al.} 2024, \aap, 684, A170

\bibitem[{{Deb} \& {Singh}(2009)}]{deb2009}
{Deb}, S. \& {Singh}, H.~P. 2009, A\&A, 507, 1729

\bibitem[{{Deupree}(1977)}]{deupree1977a}
{Deupree}, R.~G. 1977, ApJ, 211, 509

\bibitem[{{Di Criscienzo} {et~al.}(2007){Di Criscienzo}, {Caputo}, {Marconi},
  \& {Cassisi}}]{criscienzo2007}
{Di Criscienzo}, M., {Caputo}, F., {Marconi}, M., \& {Cassisi}, S. 2007, A\&A,
  471, 893

\bibitem[{{Di Criscienzo} {et~al.}(2004){Di Criscienzo}, {Marconi}, \&
  {Caputo}}]{criscienzo2004}
{Di Criscienzo}, M., {Marconi}, M., \& {Caputo}, F. 2004, ApJ, 612, 1092

\bibitem[{{Di Fabrizio} {et~al.}(2002){Di Fabrizio}, {Clementini}, {Marconi},
  {Carretta}, {Ivans}, {Bragaglia}, {Di Tomaso}, {Merighi}, {Smith}, {Sneden},
  \& {Tosi}}]{difabrizio2002}
{Di Fabrizio}, L., {Clementini}, G., {Marconi}, M., {et~al.} 2002, MNRAS, 336,
  841

\bibitem[{{Diethelm}(1983)}]{diethelm1983}
{Diethelm}, R. 1983, \aap, 124, 108

\bibitem[{{Diethelm}(1990)}]{diethelm1990}
{Diethelm}, R. 1990, \aap, 239, 186

\bibitem[{Duoandikoetxea(2001)}]{duoandikoetxea2001}
Duoandikoetxea, J. 2001, {Fourier analysis} (American Mathematical Society,
  Providence, RI, U.S.A.)

\bibitem[{Elgar(1987)}]{elgar1987}
Elgar, S. 1987, IEEE Transactions on Acoustics, Speech, and Signal Processing,
  35, 1725

\bibitem[{{Feuchtinger} \& {Dorfi}(1997)}]{feuchtinger1997}
{Feuchtinger}, M.~U. \& {Dorfi}, E.~A. 1997, \aap, 322, 817

\bibitem[{{Fiorentino} {et~al.}(2007){Fiorentino}, {Marconi}, {Musella}, \&
  {Caputo}}]{fiorentino2007}
{Fiorentino}, G., {Marconi}, M., {Musella}, I., \& {Caputo}, F. 2007, A\&A,
  476, 863

\bibitem[{{Gaia Collaboration} {et~al.}(2016){Gaia Collaboration}, {Prusti},
  {de Bruijne}, {Brown}, {Vallenari}, {Babusiaux}, {Bailer-Jones}, {Bastian},
  {Biermann}, {Evans}, {Eyer}, {Jansen}, {Jordi}, {Klioner}, {Lammers},
  {Lindegren}, {Luri}, {Mignard}, {Milligan}, {Panem}, {Poinsignon},
  {Pourbaix}, {Randich}, {Sarri}, {Sartoretti}, {Siddiqui}, {Soubiran},
  {Valette}, {van Leeuwen}, {Walton}, {Aerts}, {Arenou}, {Cropper}, {Drimmel},
  {H{\o}g}, {Katz}, {Lattanzi}, {O'Mullane}, {Grebel}, {Holland}, {Huc},
  {Passot}, {Bramante}, {Cacciari}, {Casta{\~n}eda}, {Chaoul}, {Cheek}, {De
  Angeli}, {Fabricius}, {Guerra}, {Hern{\'a}ndez}, {Jean-Antoine-Piccolo},
  {Masana}, {Messineo}, {Mowlavi}, {Nienartowicz}, {Ord{\'o}{\~n}ez-Blanco},
  {Panuzzo}, {Portell}, {Richards}, {Riello}, {Seabroke}, {Tanga},
  {Th{\'e}venin}, {Torra}, {Els}, {Gracia-Abril}, {Comoretto},
  {Garcia-Reinaldos}, {Lock}, {Mercier}, {Altmann}, {Andrae}, {Astraatmadja},
  {Bellas-Velidis}, {Benson}, {Berthier}, {Blomme}, {Busso}, {Carry},
  {Cellino}, {Clementini}, {Cowell}, {Creevey}, {Cuypers}, {Davidson}, {De
  Ridder}, {de Torres}, {Delchambre}, {Dell'Oro}, {Ducourant}, {Fr{\'e}mat},
  {Garc{\'\i}a-Torres}, {Gosset}, {Halbwachs}, {Hambly}, {Harrison}, {Hauser},
  {Hestroffer}, {Hodgkin}, {Huckle}, {Hutton}, {Jasniewicz}, {Jordan},
  {Kontizas}, {Korn}, {Lanzafame}, {Manteiga}, {Moitinho}, {Muinonen},
  {Osinde}, {Pancino}, {Pauwels}, {Petit}, {Recio-Blanco}, {Robin}, {Sarro},
  {Siopis}, {Smith}, {Smith}, {Sozzetti}, {Thuillot}, {van Reeven}, {Viala},
  {Abbas}, {Abreu Aramburu}, {Accart}, {Aguado}, {Allan}, {Allasia},
  {Altavilla}, {{\'A}lvarez}, {Alves}, {Anderson}, {Andrei}, {Anglada Varela},
  {Antiche}, {Antoja}, {Ant{\'o}n}, {Arcay}, {Atzei}, {Ayache}, {Bach},
  {Baker}, {Balaguer-N{\'u}{\~n}ez}, {Barache}, {Barata}, {Barbier}, {Barblan},
  {Baroni}, {Barrado y Navascu{\'e}s}, {Barros}, {Barstow}, {Becciani},
  {Bellazzini}, {Bellei}, {Bello Garc{\'\i}a}, {Belokurov}, {Bendjoya},
  {Berihuete}, {Bianchi}, {Bienaym{\'e}}, {Billebaud}, {Blagorodnova},
  {Blanco-Cuaresma}, {Boch}, {Bombrun}, {Borrachero}, {Bouquillon}, {Bourda},
  {Bouy}, {Bragaglia}, {Breddels}, {Brouillet}, {Br{\"u}semeister},
  {Bucciarelli}, {Budnik}, {Burgess}, {Burgon}, {Burlacu}, {Busonero}, {Buzzi},
  {Caffau}, {Cambras}, {Campbell}, {Cancelliere}, {Cantat-Gaudin}, {Carlucci},
  {Carrasco}, {Castellani}, {Charlot}, {Charnas}, {Charvet}, {Chassat},
  {Chiavassa}, {Clotet}, {Cocozza}, {Collins}, {Collins}, {Costigan}, {Crifo},
  {Cross}, {Crosta}, {Crowley}, {Dafonte}, {Damerdji}, {Dapergolas}, {David},
  {David}, {De Cat}, {de Felice}, {de Laverny}, {De Luise}, {De March}, {de
  Martino}, {de Souza}, {Debosscher}, {del Pozo}, {Delbo}, {Delgado},
  {Delgado}, {di Marco}, {Di Matteo}, {Diakite}, {Distefano}, {Dolding}, {Dos
  Anjos}, {Drazinos}, {Dur{\'a}n}, {Dzigan}, {Ecale}, {Edvardsson}, {Enke},
  {Erdmann}, {Escolar}, {Espina}, {Evans}, {Eynard Bontemps}, {Fabre},
  {Fabrizio}, {Faigler}, {Falc{\~a}o}, {Farr{\`a}s Casas}, {Faye}, {Federici},
  {Fedorets}, {Fern{\'a}ndez-Hern{\'a}ndez}, {Fernique}, {Fienga}, {Figueras},
  {Filippi}, {Findeisen}, {Fonti}, {Fouesneau}, {Fraile}, {Fraser}, {Fuchs},
  {Furnell}, {Gai}, {Galleti}, {Galluccio}, {Garabato}, {Garc{\'\i}a-Sedano},
  {Gar{\'e}}, {Garofalo}, {Garralda}, {Gavras}, {Gerssen}, {Geyer}, {Gilmore},
  {Girona}, {Giuffrida}, {Gomes}, {Gonz{\'a}lez-Marcos},
  {Gonz{\'a}lez-N{\'u}{\~n}ez}, {Gonz{\'a}lez-Vidal}, {Granvik}, {Guerrier},
  {Guillout}, {Guiraud}, {G{\'u}rpide}, {Guti{\'e}rrez-S{\'a}nchez}, {Guy},
  {Haigron}, {Hatzidimitriou}, {Haywood}, {Heiter}, {Helmi}, {Hobbs},
  {Hofmann}, {Holl}, {Holland}, {Hunt}, {Hypki}, {Icardi}, {Irwin}, {Jevardat
  de Fombelle}, {Jofr{\'e}}, {Jonker}, {Jorissen}, {Julbe}, {Karampelas},
  {Kochoska}, {Kohley}, {Kolenberg}, {Kontizas}, {Koposov}, {Kordopatis},
  {Koubsky}, {Kowalczyk}, {Krone-Martins}, {Kudryashova}, {Kull}, {Bachchan},
  {Lacoste-Seris}, {Lanza}, {Lavigne}, {Le Poncin-Lafitte}, {Lebreton},
  {Lebzelter}, {Leccia}, {Leclerc}, {Lecoeur-Taibi}, {Lemaitre}, {Lenhardt},
  {Leroux}, {Liao}, {Licata}, {Lindstr{\o}m}, {Lister}, {Livanou}, {Lobel},
  {L{\"o}ffler}, {L{\'o}pez}, {Lopez-Lozano}, {Lorenz}, {Loureiro},
  {MacDonald}, {Magalh{\~a}es Fernandes}, {Managau}, {Mann}, {Mantelet},
  {Marchal}, {Marchant}, {Marconi}, {Marie}, {Marinoni}, {Marrese},
  {Marschalk{\'o}}, {Marshall}, {Mart{\'\i}n-Fleitas}, {Martino}, {Mary},
  {Matijevi{\v{c}}}, {Mazeh}, {McMillan}, {Messina}, {Mestre}, {Michalik},
  {Millar}, {Miranda}, {Molina}, {Molinaro}, {Molinaro}, {Moln{\'a}r},
  {Moniez}, {Montegriffo}, {Monteiro}, {Mor}, {Mora}, {Morbidelli}, {Morel},
  {Morgenthaler}, {Morley}, {Morris}, {Mulone}, {Muraveva}, {Musella},
  {Narbonne}, {Nelemans}, {Nicastro}, {Noval}, {Ord{\'e}novic},
  {Ordieres-Mer{\'e}}, {Osborne}, {Pagani}, {Pagano}, {Pailler}, {Palacin},
  {Palaversa}, {Parsons}, {Paulsen}, {Pecoraro}, {Pedrosa}, {Pentik{\"a}inen},
  {Pereira}, {Pichon}, {Piersimoni}, {Pineau}, {Plachy}, {Plum}, {Poujoulet},
  {Pr{\v{s}}a}, {Pulone}, {Ragaini}, {Rago}, {Rambaux}, {Ramos-Lerate},
  {Ranalli}, {Rauw}, {Read}, {Regibo}, {Renk}, {Reyl{\'e}}, {Ribeiro},
  {Rimoldini}, {Ripepi}, {Riva}, {Rixon}, {Roelens}, {Romero-G{\'o}mez},
  {Rowell}, {Royer}, {Rudolph}, {Ruiz-Dern}, {Sadowski}, {Sagrist{\`a}
  Sell{\'e}s}, {Sahlmann}, {Salgado}, {Salguero}, {Sarasso}, {Savietto},
  {Schnorhk}, {Schultheis}, {Sciacca}, {Segol}, {Segovia}, {Segransan},
  {Serpell}, {Shih}, {Smareglia}, {Smart}, {Smith}, {Solano}, {Solitro},
  {Sordo}, {Soria Nieto}, {Souchay}, {Spagna}, {Spoto}, {Stampa}, {Steele},
  {Steidelm{\"u}ller}, {Stephenson}, {Stoev}, {Suess}, {S{\"u}veges}, {Surdej},
  {Szabados}, {Szegedi-Elek}, {Tapiador}, {Taris}, {Tauran}, {Taylor},
  {Teixeira}, {Terrett}, {Tingley}, {Trager}, {Turon}, {Ulla}, {Utrilla},
  {Valentini}, {van Elteren}, {Van Hemelryck}, {van Leeuwen}, {Varadi},
  {Vecchiato}, {Veljanoski}, {Via}, {Vicente}, {Vogt}, {Voss}, {Votruba},
  {Voutsinas}, {Walmsley}, {Weiler}, {Weingrill}, {Werner}, {Wevers},
  {Whitehead}, {Wyrzykowski}, {Yoldas}, {{\v{Z}}erjal}, {Zucker}, {Zurbach},
  {Zwitter}, {Alecu}, {Allen}, {Allende Prieto}, {Amorim},
  {Anglada-Escud{\'e}}, {Arsenijevic}, {Azaz}, {Balm}, {Beck}, {Bernstein},
  {Bigot}, {Bijaoui}, {Blasco}, {Bonfigli}, {Bono}, {Boudreault}, {Bressan},
  {Brown}, {Brunet}, {Bunclark}, {Buonanno}, {Butkevich}, {Carret}, {Carrion},
  {Chemin}, {Ch{\'e}reau}, {Corcione}, {Darmigny}, {de Boer}, {de Teodoro}, {de
  Zeeuw}, {Delle Luche}, {Domingues}, {Dubath}, {Fodor}, {Fr{\'e}zouls},
  {Fries}, {Fustes}, {Fyfe}, {Gallardo}, {Gallegos}, {Gardiol}, {Gebran},
  {Gomboc}, {G{\'o}mez}, {Grux}, {Gueguen}, {Heyrovsky}, {Hoar}, {Iannicola},
  {Isasi Parache}, {Janotto}, {Joliet}, {Jonckheere}, {Keil}, {Kim},
  {Klagyivik}, {Klar}, {Knude}, {Kochukhov}, {Kolka}, {Kos}, {Kutka}, {Lainey},
  {LeBouquin}, {Liu}, {Loreggia}, {Makarov}, {Marseille}, {Martayan},
  {Martinez-Rubi}, {Massart}, {Meynadier}, {Mignot}, {Munari}, {Nguyen},
  {Nordlander}, {Ocvirk}, {O'Flaherty}, {Olias Sanz}, {Ortiz}, {Osorio},
  {Oszkiewicz}, {Ouzounis}, {Palmer}, {Park}, {Pasquato}, {Peltzer}, {Peralta},
  {P{\'e}turaud}, {Pieniluoma}, {Pigozzi}, {Poels}, {Prat}, {Prod'homme},
  {Raison}, {Rebordao}, {Risquez}, {Rocca-Volmerange}, {Rosen}, {Ruiz-Fuertes},
  {Russo}, {Sembay}, {Serraller Vizcaino}, {Short}, {Siebert}, {Silva},
  {Sinachopoulos}, {Slezak}, {Soffel}, {Sosnowska}, {Strai{\v{z}}ys}, {ter
  Linden}, {Terrell}, {Theil}, {Tiede}, {Troisi}, {Tsalmantza}, {Tur},
  {Vaccari}, {Vachier}, {Valles}, {Van Hamme}, {Veltz}, {Virtanen}, {Wallut},
  {Wichmann}, {Wilkinson}, {Ziaeepour}, \& {Zschocke}}]{prusti2016}
{Gaia Collaboration}, {Prusti}, T., {de Bruijne}, J.~H.~J., {et~al.} 2016,
  A\&A, 595, A1

\bibitem[{{Gaia Collaboration} {et~al.}(2023){Gaia Collaboration}, {Vallenari},
  {Brown}, {Prusti}, {de Bruijne}, {Arenou}, {Babusiaux}, {Biermann},
  {Creevey}, {Ducourant}, {Evans}, {Eyer}, {Guerra}, {Hutton}, {Jordi},
  {Klioner}, {Lammers}, {Lindegren}, {Luri}, {Mignard}, {Panem}, {Pourbaix},
  {Randich}, {Sartoretti}, {Soubiran}, {Tanga}, {Walton}, {Bailer-Jones},
  {Bastian}, {Drimmel}, {Jansen}, {Katz}, {Lattanzi}, {van Leeuwen}, {Bakker},
  {Cacciari}, {Casta{\~n}eda}, {De Angeli}, {Fabricius}, {Fouesneau},
  {Fr{\'e}mat}, {Galluccio}, {Guerrier}, {Heiter}, {Masana}, {Messineo},
  {Mowlavi}, {Nicolas}, {Nienartowicz}, {Pailler}, {Panuzzo}, {Riclet}, {Roux},
  {Seabroke}, {Sordo}, {Th{\'e}venin}, {Gracia-Abril}, {Portell}, {Teyssier},
  {Altmann}, {Andrae}, {Audard}, {Bellas-Velidis}, {Benson}, {Berthier},
  {Blomme}, {Burgess}, {Busonero}, {Busso}, {C{\'a}novas}, {Carry}, {Cellino},
  {Cheek}, {Clementini}, {Damerdji}, {Davidson}, {de Teodoro}, {Nu{\~n}ez
  Campos}, {Delchambre}, {Dell'Oro}, {Esquej}, {Fern{\'a}ndez-Hern{\'a}ndez},
  {Fraile}, {Garabato}, {Garc{\'\i}a-Lario}, {Gosset}, {Haigron}, {Halbwachs},
  {Hambly}, {Harrison}, {Hern{\'a}ndez}, {Hestroffer}, {Hodgkin}, {Holl},
  {Jan{\ss}en}, {Jevardat de Fombelle}, {Jordan}, {Krone-Martins}, {Lanzafame},
  {L{\"o}ffler}, {Marchal}, {Marrese}, {Moitinho}, {Muinonen}, {Osborne},
  {Pancino}, {Pauwels}, {Recio-Blanco}, {Reyl{\'e}}, {Riello}, {Rimoldini},
  {Roegiers}, {Rybizki}, {Sarro}, {Siopis}, {Smith}, {Sozzetti}, {Utrilla},
  {van Leeuwen}, {Abbas}, {{\'A}brah{\'a}m}, {Abreu Aramburu}, {Aerts},
  {Aguado}, {Ajaj}, {Aldea-Montero}, {Altavilla}, {{\'A}lvarez}, {Alves},
  {Anders}, {Anderson}, {Anglada Varela}, {Antoja}, {Baines}, {Baker},
  {Balaguer-N{\'u}{\~n}ez}, {Balbinot}, {Balog}, {Barache}, {Barbato},
  {Barros}, {Barstow}, {Bartolom{\'e}}, {Bassilana}, {Bauchet}, {Becciani},
  {Bellazzini}, {Berihuete}, {Bernet}, {Bertone}, {Bianchi}, {Binnenfeld},
  {Blanco-Cuaresma}, {Blazere}, {Boch}, {Bombrun}, {Bossini}, {Bouquillon},
  {Bragaglia}, {Bramante}, {Breedt}, {Bressan}, {Brouillet}, {Brugaletta},
  {Bucciarelli}, {Burlacu}, {Butkevich}, {Buzzi}, {Caffau}, {Cancelliere},
  {Cantat-Gaudin}, {Carballo}, {Carlucci}, {Carnerero}, {Carrasco},
  {Casamiquela}, {Castellani}, {Castro-Ginard}, {Chaoul}, {Charlot}, {Chemin},
  {Chiaramida}, {Chiavassa}, {Chornay}, {Comoretto}, {Contursi}, {Cooper},
  {Cornez}, {Cowell}, {Crifo}, {Cropper}, {Crosta}, {Crowley}, {Dafonte},
  {Dapergolas}, {David}, {David}, {de Laverny}, {De Luise}, {De March}, {De
  Ridder}, {de Souza}, {de Torres}, {del Peloso}, {del Pozo}, {Delbo},
  {Delgado}, {Delisle}, {Demouchy}, {Dharmawardena}, {Di Matteo}, {Diakite},
  {Diener}, {Distefano}, {Dolding}, {Edvardsson}, {Enke}, {Fabre}, {Fabrizio},
  {Faigler}, {Fedorets}, {Fernique}, {Fienga}, {Figueras}, {Fournier},
  {Fouron}, {Fragkoudi}, {Gai}, {Garcia-Gutierrez}, {Garcia-Reinaldos},
  {Garc{\'\i}a-Torres}, {Garofalo}, {Gavel}, {Gavras}, {Gerlach}, {Geyer},
  {Giacobbe}, {Gilmore}, {Girona}, {Giuffrida}, {Gomel}, {Gomez},
  {Gonz{\'a}lez-N{\'u}{\~n}ez}, {Gonz{\'a}lez-Santamar{\'\i}a},
  {Gonz{\'a}lez-Vidal}, {Granvik}, {Guillout}, {Guiraud},
  {Guti{\'e}rrez-S{\'a}nchez}, {Guy}, {Hatzidimitriou}, {Hauser}, {Haywood},
  {Helmer}, {Helmi}, {Sarmiento}, {Hidalgo}, {Hilger}, {H{\l}adczuk}, {Hobbs},
  {Holland}, {Huckle}, {Jardine}, {Jasniewicz}, {Jean-Antoine Piccolo},
  {Jim{\'e}nez-Arranz}, {Jorissen}, {Juaristi Campillo}, {Julbe}, {Karbevska},
  {Kervella}, {Khanna}, {Kontizas}, {Kordopatis}, {Korn}, {K{\'o}sp{\'a}l},
  {Kostrzewa-Rutkowska}, {Kruszy{\'n}ska}, {Kun}, {Laizeau}, {Lambert},
  {Lanza}, {Lasne}, {Le Campion}, {Lebreton}, {Lebzelter}, {Leccia}, {Leclerc},
  {Lecoeur-Taibi}, {Liao}, {Licata}, {Lindstr{\o}m}, {Lister}, {Livanou},
  {Lobel}, {Lorca}, {Loup}, {Madrero Pardo}, {Magdaleno Romeo}, {Managau},
  {Mann}, {Manteiga}, {Marchant}, {Marconi}, {Marcos}, {Marcos Santos},
  {Mar{\'\i}n Pina}, {Marinoni}, {Marocco}, {Marshall}, {Martin Polo},
  {Mart{\'\i}n-Fleitas}, {Marton}, {Mary}, {Masip}, {Massari},
  {Mastrobuono-Battisti}, {Mazeh}, {McMillan}, {Messina}, {Michalik}, {Millar},
  {Mints}, {Molina}, {Molinaro}, {Moln{\'a}r}, {Monari}, {Mongui{\'o}},
  {Montegriffo}, {Montero}, {Mor}, {Mora}, {Morbidelli}, {Morel}, {Morris},
  {Muraveva}, {Murphy}, {Musella}, {Nagy}, {Noval}, {Oca{\~n}a}, {Ogden},
  {Ordenovic}, {Osinde}, {Pagani}, {Pagano}, {Palaversa}, {Palicio},
  {Pallas-Quintela}, {Panahi}, {Payne-Wardenaar}, {Pe{\~n}alosa Esteller},
  {Penttil{\"a}}, {Pichon}, {Piersimoni}, {Pineau}, {Plachy}, {Plum}, {Poggio},
  {Pr{\v{s}}a}, {Pulone}, {Racero}, {Ragaini}, {Rainer}, {Raiteri}, {Rambaux},
  {Ramos}, {Ramos-Lerate}, {Re Fiorentin}, {Regibo}, {Richards}, {Rios Diaz},
  {Ripepi}, {Riva}, {Rix}, {Rixon}, {Robichon}, {Robin}, {Robin}, {Roelens},
  {Rogues}, {Rohrbasser}, {Romero-G{\'o}mez}, {Rowell}, {Royer}, {Ruz Mieres},
  {Rybicki}, {Sadowski}, {S{\'a}ez N{\'u}{\~n}ez}, {Sagrist{\`a} Sell{\'e}s},
  {Sahlmann}, {Salguero}, {Samaras}, {Sanchez Gimenez}, {Sanna},
  {Santove{\~n}a}, {Sarasso}, {Schultheis}, {Sciacca}, {Segol}, {Segovia},
  {S{\'e}gransan}, {Semeux}, {Shahaf}, {Siddiqui}, {Siebert}, {Siltala},
  {Silvelo}, {Slezak}, {Slezak}, {Smart}, {Snaith}, {Solano}, {Solitro},
  {Souami}, {Souchay}, {Spagna}, {Spina}, {Spoto}, {Steele},
  {Steidelm{\"u}ller}, {Stephenson}, {S{\"u}veges}, {Surdej}, {Szabados},
  {Szegedi-Elek}, {Taris}, {Taylor}, {Teixeira}, {Tolomei}, {Tonello}, {Torra},
  {Torra}, {Torralba Elipe}, {Trabucchi}, {Tsounis}, {Turon}, {Ulla}, {Unger},
  {Vaillant}, {van Dillen}, {van Reeven}, {Vanel}, {Vecchiato}, {Viala},
  {Vicente}, {Voutsinas}, {Weiler}, {Wevers}, {Wyrzykowski}, {Yoldas}, {Yvard},
  {Zhao}, {Zorec}, {Zucker}, \& {Zwitter}}]{vallenari2023}
{Gaia Collaboration}, {Vallenari}, A., {Brown}, A.~G.~A., {et~al.} 2023, \aap,
  674, A1

\bibitem[{{Gao} {et~al.}(2013){Gao}, {Jiang}, {Li}, \& {Xue}}]{gao2013}
{Gao}, J., {Jiang}, B.~W., {Li}, A., \& {Xue}, M.~Y. 2013, \apj, 776, 7

\bibitem[{{Gingold}(1985)}]{gingold1985}
{Gingold}, R.~A. 1985, MmSAI, 56, 169

\bibitem[{{Gordon} {et~al.}(2003){Gordon}, {Clayton}, {Misselt}, {Landolt}, \&
  {Wolff}}]{gordon2003}
{Gordon}, K.~D., {Clayton}, G.~C., {Misselt}, K.~A., {Landolt}, A.~U., \&
  {Wolff}, M.~J. 2003, \apj, 594, 279

\bibitem[{{Groenewegen} \& {Jurkovic}(2017)}]{groenewegen2017b}
{Groenewegen}, M.~A.~T. \& {Jurkovic}, M.~I. 2017, A\&A, 604, A29

\bibitem[{{Hinshaw} {et~al.}(2013){Hinshaw}, {Larson}, {Komatsu}, {Spergel},
  {Bennett}, {Dunkley}, {Nolta}, {Halpern}, {Hill}, {Odegard}, {Page}, {Smith},
  {Weiland}, {Gold}, {Jarosik}, {Kogut}, {Limon}, {Meyer}, {Tucker}, {Wollack},
  \& {Wright}}]{hinshaw2013}
{Hinshaw}, G., {Larson}, D., {Komatsu}, E., {et~al.} 2013, ApJS, 208, 19

\bibitem[{{Jermyn} {et~al.}(2023){Jermyn}, {Bauer}, {Schwab}, {Farmer}, {Ball},
  {Bellinger}, {Dotter}, {Joyce}, {Marchant}, {Mombarg}, {Wolf}, {Sunny Wong},
  {Cinquegrana}, {Farrell}, {Smolec}, {Thoul}, {Cantiello}, {Herwig}, {Toloza},
  {Bildsten}, {Townsend}, \& {Timmes}}]{jermyn2023}
{Jermyn}, A.~S., {Bauer}, E.~B., {Schwab}, J., {et~al.} 2023, \apjs, 265, 15

\bibitem[{{Joyce} {et~al.}(2023){Joyce}, {Johnson}, {Marchetti}, {Rich},
  {Simion}, \& {Bourke}}]{joyce2023}
{Joyce}, M., {Johnson}, C.~I., {Marchetti}, T., {et~al.} 2023, ApJ, 946, 28

\bibitem[{{Jurcsik} \& {Kovacs}(1996)}]{jurcsik1996}
{Jurcsik}, J. \& {Kovacs}, G. 1996, A\&A, 312, 111

\bibitem[{{Keller} \& {Wood}(2002)}]{keller2002}
{Keller}, S.~C. \& {Wood}, P.~R. 2002, \apj, 578, 144

\bibitem[{{Keller} \& {Wood}(2006)}]{keller2006}
{Keller}, S.~C. \& {Wood}, P.~R. 2006, \apj, 642, 834

\bibitem[{{Koll{\'a}th} {et~al.}(2011){Koll{\'a}th}, {Moln{\'a}r}, \&
  {Szab{\'o}}}]{kollath-2011}
{Koll{\'a}th}, Z., {Moln{\'a}r}, L., \& {Szab{\'o}}, R. 2011, \mnras, 414, 1111

\bibitem[{Kolmogorov(1933)}]{kolmogorov1933}
Kolmogorov, A. 1933, Giornale dell' Istituto Italiano degli Attuari, 4, 83

\bibitem[{{Kovacs} \& {Kanbur}(1998)}]{kovacs1998}
{Kovacs}, G. \& {Kanbur}, S.~M. 1998, MNRAS, 295, 834

\bibitem[{{Kov{\'a}cs} {et~al.}(2023){Kov{\'a}cs}, {Nuspl}, \&
  {Szab{\'o}}}]{kovacs2023}
{Kov{\'a}cs}, G.~B., {Nuspl}, J., \& {Szab{\'o}}, R. 2023, MNRAS, 521, 4878

\bibitem[{{Kov{\'a}cs} {et~al.}(2024){Kov{\'a}cs}, {Nuspl}, \&
  {Szab{\'o}}}]{kovacs2024}
{Kov{\'a}cs}, G.~B., {Nuspl}, J., \& {Szab{\'o}}, R. 2024, \mnras, 527, L1

\bibitem[{{Kumar} {et~al.}(2024){Kumar}, {Bhardwaj}, {Singh}, {Rejkuba},
  {Marconi}, \& {Prugniel}}]{kumar2024}
{Kumar}, N., {Bhardwaj}, A., {Singh}, H.~P., {et~al.} 2024, \mnras, 531, 2976

\bibitem[{{Madore}(1982)}]{madore1982}
{Madore}, B.~F. 1982, ApJ, 253, 575

\bibitem[{{Majaess}(2010)}]{majaess2010}
{Majaess}, D.~J. 2010, Journal of the American Association of Variable Star
  Observers (JAAVSO), 38, 100

\bibitem[{{Marconi}(2017)}]{marconi2017b}
{Marconi}, M. 2017, in European Physical Journal Web of Conferences, Vol. 152,
  European Physical Journal Web of Conferences, 06001

\bibitem[{{Marconi} \& {Clementini}(2005)}]{marconi2005}
{Marconi}, M. \& {Clementini}, G. 2005, AJ, 129, 2257

\bibitem[{{Marconi} {et~al.}(2024){Marconi}, {De Somma}, {Molinaro},
  {Bhardwaj}, {Ripepi}, {Musella}, {Sicignano}, {Trentin}, \&
  {Leccia}}]{marconi2024}
{Marconi}, M., {De Somma}, G., {Molinaro}, R., {et~al.} 2024, \mnras, 529, 4210

\bibitem[{{Marconi} \& {Degl'Innocenti}(2007)}]{marconi2007b}
{Marconi}, M. \& {Degl'Innocenti}, S. 2007, \aap, 474, 557

\bibitem[{{Marconi} \& {Di Criscienzo}(2007)}]{marconi2007a}
{Marconi}, M. \& {Di Criscienzo}, M. 2007, A\&A, 467, 223

\bibitem[{{Marconi} {et~al.}(2013{\natexlab{a}}){Marconi}, {Molinaro}, {Bono},
  {Pietrzy{\'n}ski}, {Gieren}, {Pilecki}, {Stellingwerf}, {Graczyk}, {Smolec},
  {Konorski}, {Suchomska}, {G{\'o}rski}, \& {Karczmarek}}]{marconi2013b}
{Marconi}, M., {Molinaro}, R., {Bono}, G., {et~al.} 2013{\natexlab{a}}, ApJL,
  768, L6

\bibitem[{{Marconi} {et~al.}(2017){Marconi}, {Molinaro}, {Ripepi}, {Cioni},
  {Clementini}, {Moretti}, {Ragosta}, {de Grijs}, {Groenewegen}, \&
  {Ivanov}}]{marconi2017a}
{Marconi}, M., {Molinaro}, R., {Ripepi}, V., {et~al.} 2017, MNRAS, 466, 3206

\bibitem[{{Marconi} {et~al.}(2013{\natexlab{b}}){Marconi}, {Molinaro},
  {Ripepi}, {Musella}, \& {Brocato}}]{marconi2013a}
{Marconi}, M., {Molinaro}, R., {Ripepi}, V., {Musella}, I., \& {Brocato}, E.
  2013{\natexlab{b}}, MNRAS, 428, 2185

\bibitem[{{Matsunaga} {et~al.}(2009){Matsunaga}, {Feast}, \&
  {Menzies}}]{matsunaga2009}
{Matsunaga}, N., {Feast}, M.~W., \& {Menzies}, J.~W. 2009, MNRAS, 397, 933

\bibitem[{{Matsunaga} {et~al.}(2011){Matsunaga}, {Feast}, \&
  {Soszy{\'n}ski}}]{matsunaga2011}
{Matsunaga}, N., {Feast}, M.~W., \& {Soszy{\'n}ski}, I. 2011, MNRAS, 413, 223

\bibitem[{{Matsunaga} {et~al.}(2006){Matsunaga}, {Fukushi}, {Nakada},
  {Tanab{\'e}}, {Feast}, {Menzies}, {Ita}, {Nishiyama}, {Baba}, {Naoi},
  {Nakaya}, {Kawadu}, {Ishihara}, \& {Kato}}]{matsunaga2006}
{Matsunaga}, N., {Fukushi}, H., {Nakada}, Y., {et~al.} 2006, MNRAS, 370, 1979

\bibitem[{{Moln{\'a}r} {et~al.}(2012){Moln{\'a}r}, {Koll{\'a}th}, {Szab{\'o}},
  {Bryson}, {Kolenberg}, {Mullally}, \& {Thompson}}]{molnar-2012}
{Moln{\'a}r}, L., {Koll{\'a}th}, Z., {Szab{\'o}}, R., {et~al.} 2012, \apjl,
  757, L13

\bibitem[{{Moskalik} \& {Buchler}(1993)}]{moskalik1993}
{Moskalik}, P. \& {Buchler}, J.~R. 1993, \apj, 406, 190

\bibitem[{{Nemec} {et~al.}(2013){Nemec}, {Cohen}, {Ripepi}, {Derekas},
  {Moskalik}, {Sesar}, {Chadid}, \& {Bruntt}}]{nemec2013}
{Nemec}, J.~M., {Cohen}, J.~G., {Ripepi}, V., {et~al.} 2013, ApJ, 773, 181

\bibitem[{{Netzel} {et~al.}(2023){Netzel}, {Moln{\'a}r}, \&
  {Joyce}}]{netzel2023}
{Netzel}, H., {Moln{\'a}r}, L., \& {Joyce}, M. 2023, MNRAS, 525, 5378

\bibitem[{{Ngeow} {et~al.}(2022){Ngeow}, {Bhardwaj}, {Henderson}, {Graham},
  {Laher}, {Medford}, {Purdum}, \& {Rusholme}}]{ngeow2022}
{Ngeow}, C.-C., {Bhardwaj}, A., {Henderson}, J.-Y., {et~al.} 2022, AJ, 164, 154

\bibitem[{{Paxton} {et~al.}(2011){Paxton}, {Bildsten}, {Dotter}, {Herwig},
  {Lesaffre}, \& {Timmes}}]{paxton2011}
{Paxton}, B., {Bildsten}, L., {Dotter}, A., {et~al.} 2011, ApJS, 192, 3

\bibitem[{{Paxton} {et~al.}(2013){Paxton}, {Cantiello}, {Arras}, {Bildsten},
  {Brown}, {Dotter}, {Mankovich}, {Montgomery}, {Stello}, {Timmes}, \&
  {Townsend}}]{paxton2013}
{Paxton}, B., {Cantiello}, M., {Arras}, P., {et~al.} 2013, ApJS, 208, 4

\bibitem[{{Paxton} {et~al.}(2015){Paxton}, {Marchant}, {Schwab}, {Bauer},
  {Bildsten}, {Cantiello}, {Dessart}, {Farmer}, {Hu}, {Langer}, {Townsend},
  {Townsley}, \& {Timmes}}]{paxton2015}
{Paxton}, B., {Marchant}, P., {Schwab}, J., {et~al.} 2015, ApJS, 220, 15

\bibitem[{{Paxton} {et~al.}(2018){Paxton}, {Schwab}, {Bauer}, {Bildsten},
  {Blinnikov}, {Duffell}, {Farmer}, {Goldberg}, {Marchant}, {Sorokina},
  {Thoul}, {Townsend}, \& {Timmes}}]{paxton2018}
{Paxton}, B., {Schwab}, J., {Bauer}, E.~B., {et~al.} 2018, ApJS, 234, 34

\bibitem[{{Paxton} {et~al.}(2019){Paxton}, {Smolec}, {Schwab}, {Gautschy},
  {Bildsten}, {Cantiello}, {Dotter}, {Farmer}, {Goldberg}, {Jermyn}, {Kanbur},
  {Marchant}, {Thoul}, {Townsend}, {Wolf}, {Zhang}, \& {Timmes}}]{paxton2019}
{Paxton}, B., {Smolec}, R., {Schwab}, J., {et~al.} 2019, ApJS, 243, 10

\bibitem[{{Petersen}(1980)}]{petersen1980}
{Petersen}, J.~O. 1980, \ssr, 27, 495

\bibitem[{{Pietrzy{\'n}ski} {et~al.}(2019){Pietrzy{\'n}ski}, {Graczyk},
  {Gallenne}, {Gieren}, {Thompson}, {Pilecki}, {Karczmarek}, {G{\'o}rski},
  {Suchomska}, {Taormina}, {Zgirski}, {Wielg{\'o}rski}, {Ko{\l}aczkowski},
  {Konorski}, {Villanova}, {Nardetto}, {Kervella}, {Bresolin}, {Kudritzki},
  {Storm}, {Smolec}, \& {Narloch}}]{pietrzynski2019}
{Pietrzy{\'n}ski}, G., {Graczyk}, D., {Gallenne}, A., {et~al.} 2019, \nat, 567,
  200

\bibitem[{{Ripepi} {et~al.}(2023){Ripepi}, {Clementini}, {Molinaro}, {Leccia},
  {Plachy}, {Moln{\'a}r}, {Rimoldini}, {Musella}, {Marconi}, {Garofalo},
  {Audard}, {Holl}, {Evans}, {Jevardat de Fombelle}, {Lecoeur-Taibi},
  {Marchal}, {Mowlavi}, {Muraveva}, {Nienartowicz}, {Sartoretti}, {Szabados},
  \& {Eyer}}]{ripepi2023}
{Ripepi}, V., {Clementini}, G., {Molinaro}, R., {et~al.} 2023, \aap, 674, A17

\bibitem[{{Ripepi} {et~al.}(2019){Ripepi}, {Molinaro}, {Musella}, {Marconi},
  {Leccia}, \& {Eyer}}]{ripepi2019}
{Ripepi}, V., {Molinaro}, R., {Musella}, I., {et~al.} 2019, A\&A, 625, A14

\bibitem[{{Sandage} {et~al.}(1994){Sandage}, {Diethelm}, \&
  {Tammann}}]{sandage1994}
{Sandage}, A., {Diethelm}, R., \& {Tammann}, G.~A. 1994, \aap, 283, 111

\bibitem[{{Schlegel} {et~al.}(1998){Schlegel}, {Finkbeiner}, \&
  {Davis}}]{schlegel1998}
{Schlegel}, D.~J., {Finkbeiner}, D.~P., \& {Davis}, M. 1998, ApJ, 500, 525

\bibitem[{{Simon} \& {Davis}(1983)}]{simon1983}
{Simon}, N.~R. \& {Davis}, C.~G. 1983, ApJ, 266, 787

\bibitem[{{Simon} \& {Lee}(1981)}]{simon1981}
{Simon}, N.~R. \& {Lee}, A.~S. 1981, ApJ, 248, 291

\bibitem[{{Simon} \& {Teays}(1982)}]{simon1982}
{Simon}, N.~R. \& {Teays}, T.~J. 1982, ApJ, 261, 586

\bibitem[{{Skowron} {et~al.}(2021){Skowron}, {Skowron}, {Udalski},
  {Szyma{\'n}ski}, {Soszy{\'n}ski}, {Wyrzykowski}, {Ulaczyk}, {Poleski},
  {Koz{\l}owski}, {Pietrukowicz}, {Mr{\'o}z}, {Rybicki}, {Iwanek}, {Wrona}, \&
  {Gromadzki}}]{skowron2021}
{Skowron}, D.~M., {Skowron}, J., {Udalski}, A., {et~al.} 2021, ApJS, 252, 23

\bibitem[{Smirnov(1948)}]{smirnov1948}
Smirnov, N. 1948, The Annals of Mathematical Statistics, 19, 279

\bibitem[{{Smolec}(2005)}]{smolec2005}
{Smolec}, R. 2005, AcA, 55, 59

\bibitem[{{Smolec}(2016)}]{smolec-2016}
{Smolec}, R. 2016, \mnras, 456, 3475

\bibitem[{{Smolec} \& {Moskalik}(2008)}]{smolec2008}
{Smolec}, R. \& {Moskalik}, P. 2008, AcA, 58, 193

\bibitem[{{Smolec} \& {Moskalik}(2012)}]{smolec-2012}
{Smolec}, R. \& {Moskalik}, P. 2012, \mnras, 426, 108

\bibitem[{{Smolec} \& {Moskalik}(2014)}]{smolec-2014}
{Smolec}, R. \& {Moskalik}, P. 2014, \mnras, 441, 101

\bibitem[{{Smolec} {et~al.}(2013){Smolec}, {Pietrzy{\'n}ski}, {Graczyk},
  {Pilecki}, {Gieren}, {Thompson}, {St{\k{e}}pie{\'n}}, {Karczmarek},
  {Konorski}, {G{\'o}rski}, {Suchomska}, {Bono}, {Prada}, \&
  {Nardetto}}]{smolec2013}
{Smolec}, R., {Pietrzy{\'n}ski}, G., {Graczyk}, D., {et~al.} 2013, MNRAS, 428,
  3034

\bibitem[{{Smolec} {et~al.}(2012){Smolec}, {Soszy{\'n}ski}, {Moskalik},
  {Udalski}, {Szyma{\'n}ski}, {Kubiak}, {Pietrzy{\'n}ski}, {Wyrzykowski},
  {Ulaczyk}, {Poleski}, {Koz{\l}owski}, \& {Pietrukowicz}}]{smolec2012a}
{Smolec}, R., {Soszy{\'n}ski}, I., {Moskalik}, P., {et~al.} 2012, MNRAS, 419,
  2407

\bibitem[{{Soszy{\'n}ski} {et~al.}(2011){Soszy{\'n}ski}, {Udalski},
  {Pietrukowicz}, {Szyma{\'n}ski}, {Kubiak}, {Pietrzy{\'n}ski}, {Wyrzykowski},
  {Ulaczyk}, {Poleski}, \& {Koz{\l}owski}}]{soszynski2011}
{Soszy{\'n}ski}, I., {Udalski}, A., {Pietrukowicz}, P., {et~al.} 2011, AcA, 61,
  285

\bibitem[{{Soszy{\'n}ski} {et~al.}(2015){Soszy{\'n}ski}, {Udalski},
  {Szyma{\'n}ski}, {Skowron}, {Pietrzy{\'n}ski}, {Poleski}, {Pietrukowicz},
  {Skowron}, {Mr{\'o}z}, {Koz{\l}owski}, {Wyrzykowski}, {Ulaczyk}, \&
  {Pawlak}}]{soszynski2015}
{Soszy{\'n}ski}, I., {Udalski}, A., {Szyma{\'n}ski}, M.~K., {et~al.} 2015, AcA,
  65, 297

\bibitem[{{Soszy{\'n}ski} {et~al.}(2018){Soszy{\'n}ski}, {Udalski},
  {Szyma{\'n}ski}, {Wyrzykowski}, {Ulaczyk}, {Poleski}, {Pietrukowicz},
  {Koz{\l}owski}, {Skowron}, {Skowron}, {Mr{\'o}z}, {Rybicki}, \&
  {Iwanek}}]{soszynski2018}
{Soszy{\'n}ski}, I., {Udalski}, A., {Szyma{\'n}ski}, M.~K., {et~al.} 2018, AcA,
  68, 89

\bibitem[{{Stellingwerf}(1982{\natexlab{a}})}]{stellingwerf1982a}
{Stellingwerf}, R.~F. 1982{\natexlab{a}}, ApJ, 262, 330

\bibitem[{{Stellingwerf}(1982{\natexlab{b}})}]{stellingwerf1982b}
{Stellingwerf}, R.~F. 1982{\natexlab{b}}, ApJ, 262, 339

\bibitem[{{Stellingwerf} \& {Donohoe}(1987)}]{stellingwerf1987}
{Stellingwerf}, R.~F. \& {Donohoe}, M. 1987, ApJ, 314, 252

\bibitem[{{Tammann} {et~al.}(2003){Tammann}, {Sandage}, \&
  {Reindl}}]{tammann2003}
{Tammann}, G.~A., {Sandage}, A., \& {Reindl}, B. 2003, A\&A, 404, 423

\bibitem[{{Wang} \& {Chen}(2023)}]{wang2023}
{Wang}, S. \& {Chen}, X. 2023, \apj, 946, 43

\bibitem[{{Wielg{\'o}rski} {et~al.}(2022){Wielg{\'o}rski}, {Pietrzy{\'n}ski},
  {Pilecki}, {Gieren}, {Zgirski}, {G{\'o}rski}, {Hajdu}, {Narloch},
  {Karczmarek}, {Smolec}, {Kervella}, {Storm}, {Gallenne}, {Breuval}, {Lewis},
  {Ka{\l}uszy{\'n}ski}, {Graczyk}, {Pych}, {Suchomska}, {Taormina}, {Rojas
  Garcia}, {Kotek}, {Chini}, {Pozo N{\~{u}}nez}, {Noroozi}, {Sobrino Figaredo},
  {Haas}, {Hodapp}, {Miko{\l}ajczyk}, {Kotysz}, {Mo{\'z}dzierski}, \&
  {Ko{\l}aczek-Szyma{\'n}ski}}]{wielgorski2022}
{Wielg{\'o}rski}, P., {Pietrzy{\'n}ski}, G., {Pilecki}, B., {et~al.} 2022, ApJ,
  927, 89

\bibitem[{{Wood} {et~al.}(1997){Wood}, {S.~Arnold}, \& {Sebo}}]{wood1997}
{Wood}, P.~R., {S.~Arnold}, A., \& {Sebo}, K.~M. 1997, ApJL, 485, L25

\end{thebibliography}

\appendix

\section{Chemical compositions of the BL~Her models}
In Table~\ref{tab:composition}, we present the equivalent $ZX$ values for each [Fe/H] value.
\begin{table}
\caption{Chemical compositions of the adopted pulsation models.}
\centering
\begin{tabular}{c c c}
\hline
[Fe/H] & $Z$ & $X$\\
\hline
-2.00 & 0.00014 & 0.75115\\
-1.50 & 0.00043 & 0.75041\\
-1.35 &0.00061& 0.74996\\
-1.00 &0.00135& 0.74806\\
-0.50&0.00424&0.74073\\
-0.20 &0.00834& 0.73032\\
0.00 &0.01300& 0.71847\\
\hline
\end{tabular}
\tablefoot{\small
	 The $Z$ and $X$ values are estimated from the $\mathrm{[Fe/H]}$ values by assuming the primordial helium value of 0.2485 from the WMAP CMB observations \citep{hinshaw2013} and the helium enrichment parameter value of 1.54 \citep{asplund2009}. The solar mixture is adopted from \citet{asplund2009}.}
\label{tab:composition}
\end{table}

\section{Additional figures}

\begin{figure*}
\centering
\includegraphics[scale = 1]{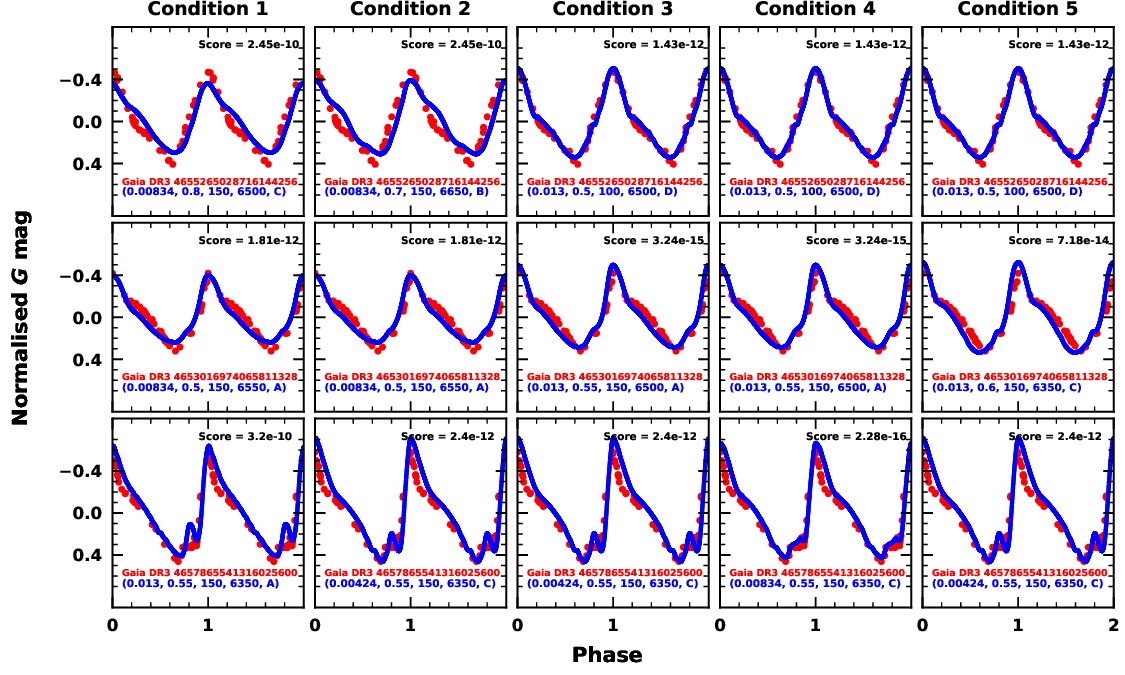}
\caption{A visualisation of the modeled-observed pairs obtained from preliminary exercises carried out using different conditions to determine the robust weights on the light curve parameters for calculating the parameter $d$ as in Eq.~\ref{eq:d}. The input stellar parameters of the corresponding models are included in the format ($Z, M/M_{\odot}, L/L_{\odot}, T_{\rm eff}$, convection set) in each sub-plot.}
\label{fig:diffN}
\end{figure*}

In Fig.~\ref{fig:diffN}, we exhibit the different modeled-observed pairs obtained from preliminary exercises carried out using different conditions to determine the robust weights on the light curve parameters for calculating the parameter $d$ as in Eq.~\ref{eq:d}. The light curve parameters are $p \in \{ \log(P), A, S_k, A_c, R_{i1}, \phi_{i1} \}$ and the different conditions used are as follows:
\begin{enumerate}
    \item Condition 1: $n_p=1$ for all
    \item Condition 2: $n_p=5$ for $A$, $S_k$; $n_p=1$ for the rest
    \item Condition 3: $n_p=10$ for $A$, $S_k$; $n_p=1$ for the rest
    \item Condition 4: $n_p=20$ for $A$, $S_k$; $n_p=1$ for the rest
    \item Condition 5: $n_p=10$ for $\log(P)$, $A$, $S_k$; $n_p=1$ for the rest
\end{enumerate}

The reason for placing higher weights on $A$ and $S_k$ (and not on the other parameters) arises from the fact that the theoretical and observed light curves in this work tended to be more ``offset'' either in terms of their amplitudes or their skewness; this is also evident from the \textit{silver sample} (Fig.~\ref{fig:LC_silver}). The reader is advised to set weights on the light curve parameters in the determination of the parameter $d$ based on their datasets.

\begin{figure*}
\centering
\includegraphics[scale = 1]{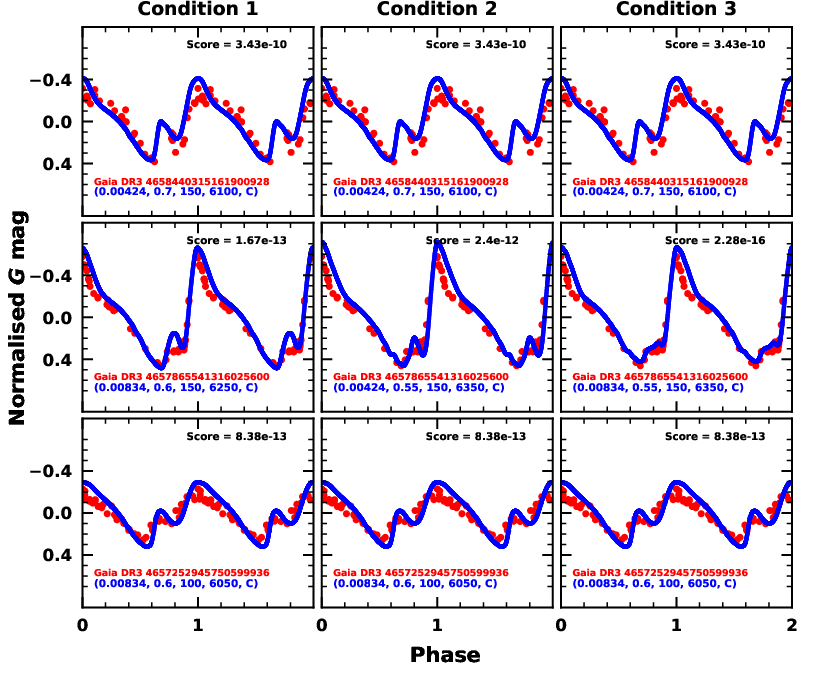}
\caption{A visualisation of the modeled-observed pairs obtained from using different conditions for the cases where the observed light curves were fitted with order of fit, $N=7$. The input stellar parameters of the corresponding models are included in the format ($Z, M/M_{\odot}, L/L_{\odot}, T_{\rm eff}$, convection set) in each sub-plot.}
\label{fig:phi}
\end{figure*}

In Fig.~\ref{fig:phi}, we present the modeled-observed pairs obtained from using different conditions for the cases where the observed light curves were fitted with order of fit, $N=7$. The different conditions used are as follows:
\begin{enumerate}
    \item Condition 1: $p \in \{ \log(P), A, S_k, A_c, R_{21}, R_{31}, \phi_{21}, \phi_{31} \}$
    \item Condition 2: $p \in \{ \log(P), A, S_k, A_c, R_{21}, R_{31}, R_{41}, R_{51}, R_{61}, R_{71}, \phi_{21}, \phi_{31} \}$
    \item Condition 3: $p \in \{ \log(P), A, S_k, A_c, R_{21}, R_{31}, R_{41}, R_{51}, R_{61}, R_{71}, \phi_{21}, \phi_{31} , \phi_{41}, \phi_{51}, \phi_{61}, \phi_{71}\}$
\end{enumerate}

\begin{figure*}[h!]
\centering
\includegraphics[scale = 1]{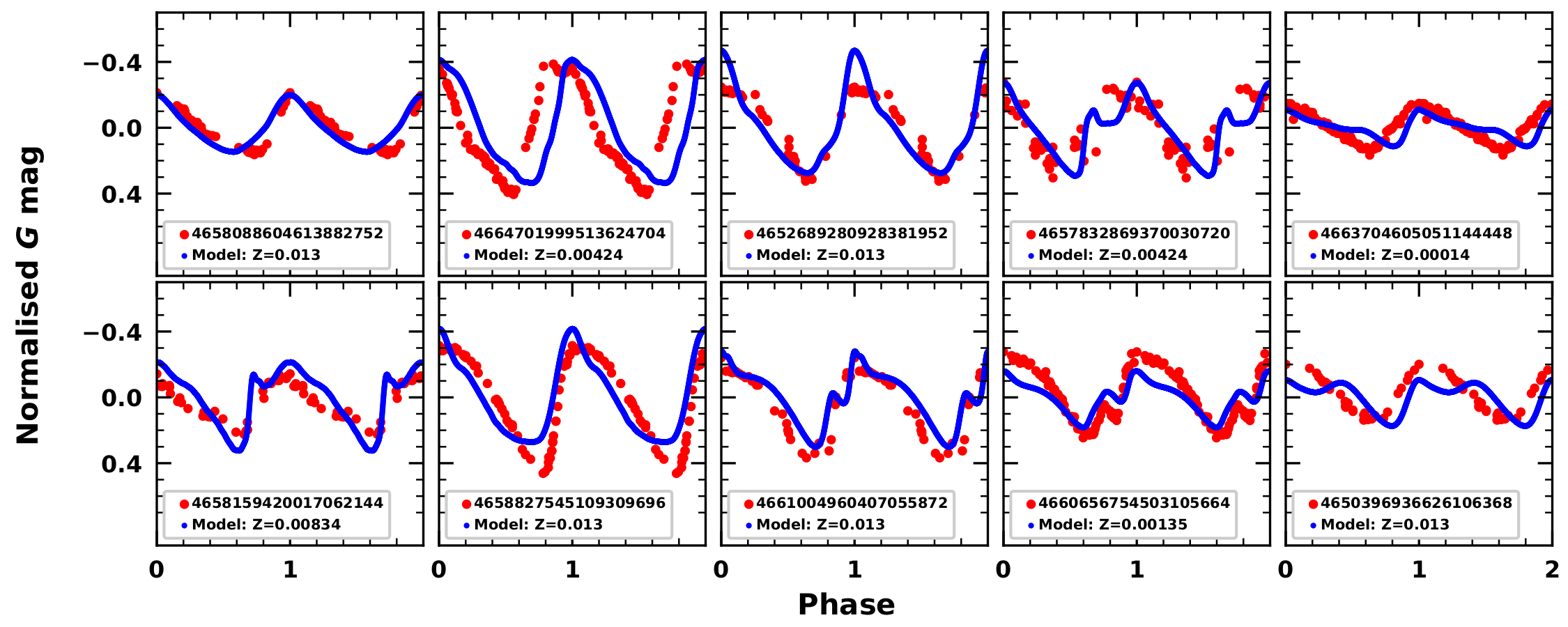}
\caption{The light curves of the rejected model (blue)-observed (red) pairs of the 10 BL~Her stars in the LMC. Note that the BL~Her star $Gaia$~DR3~4657832869370030720 has sparse epoch photometry in the $G$ band.}
\label{fig:rejected_LC}
\end{figure*}

\begin{figure*}
\centering
\includegraphics[scale = 0.95]{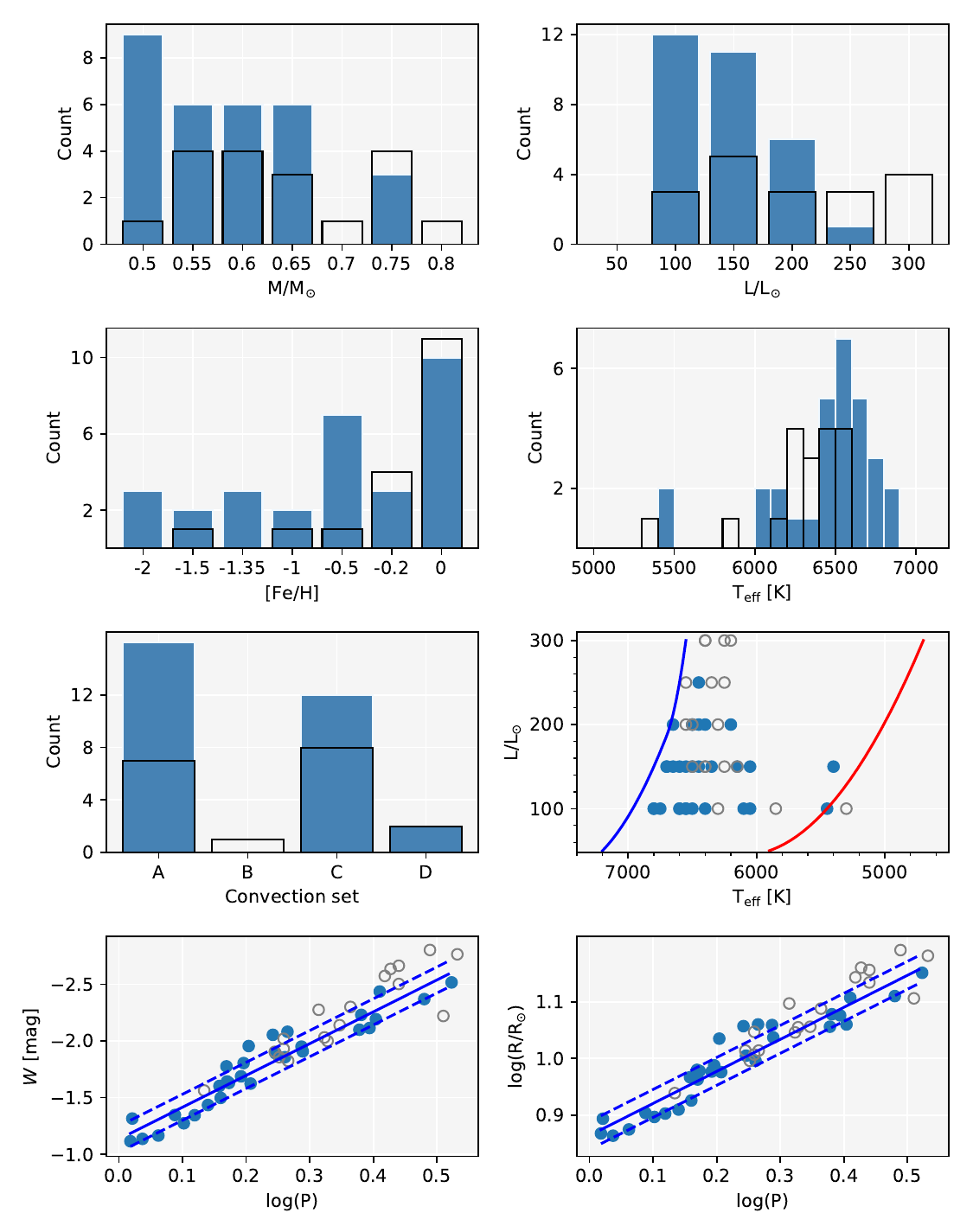}
\caption{Same as Fig.~\ref{fig:table_properties} but for both the \textit{gold sample} and the \textit{silver sample} of the BL~Her models. The open-faced histogram bars and circles represent the \textit{silver sample} BL~Her models.}
\label{fig:table_properties2}
\end{figure*}

\begin{figure*}
\centering
\includegraphics[scale = 1]{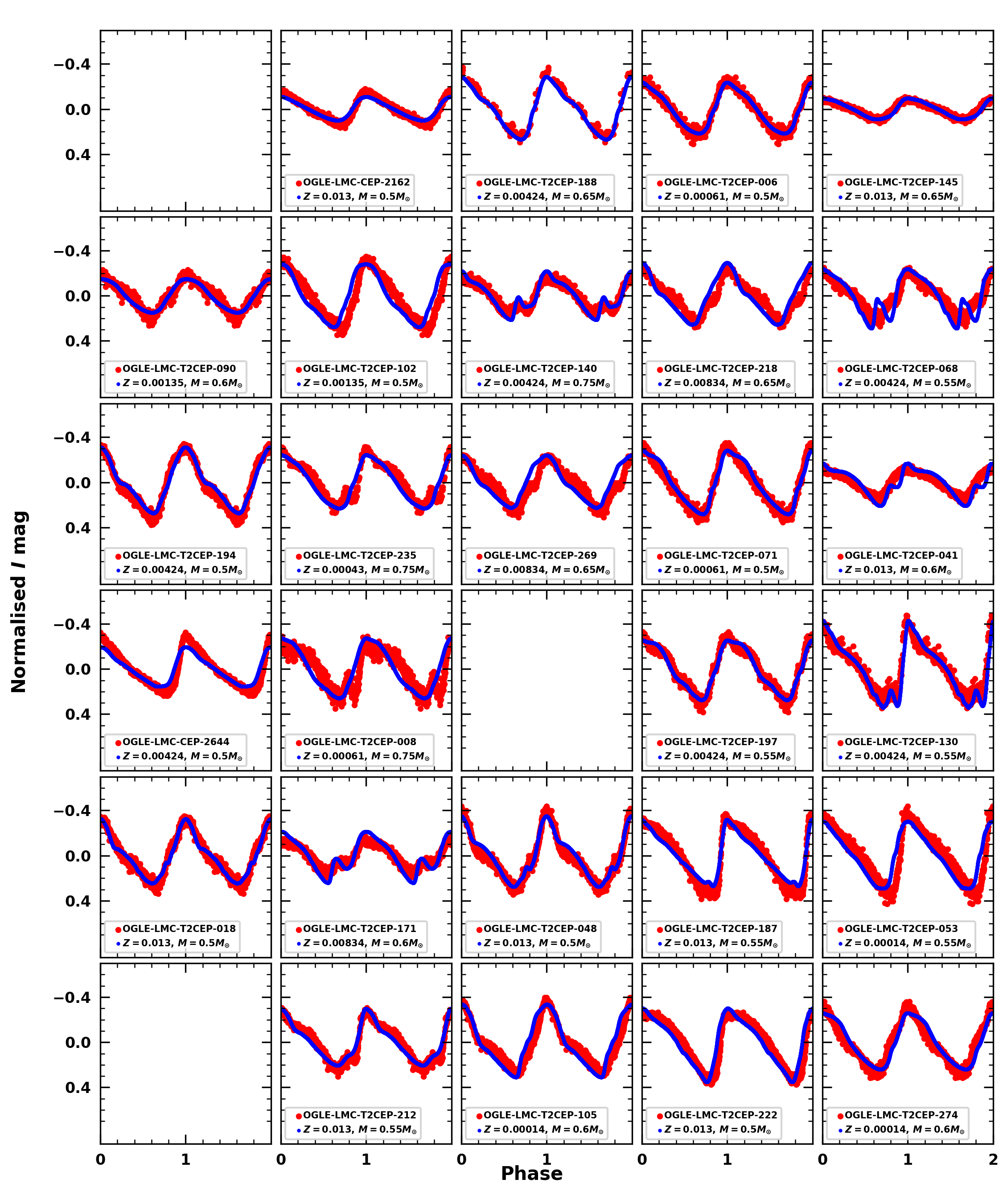}
\caption{Same as Fig.~\ref{fig:LC_gold} but for their OGLE counterparts. Missing sub-plots indicate lack of their OGLE counterparts and/or lack of $VI$ photometric data.}
\label{fig:LC_gold_ogle}
\end{figure*}

In Fig.~\ref{fig:rejected_LC}, we display the light curves of the 10 modeled-observed pairs of BL~Her stars that were rejected from this analysis. Fig.~\ref{fig:table_properties2} shows the distribution of the subset of the \textit{gold sample} and the \textit{silver sample} of BL~Her models that match the best with BL~Her stars in the LMC as as a function of metallicity [Fe/H], stellar mass $M/M_{\odot}$, stellar luminosity $L/L_{\odot}$, effective temperature and the convection parameter sets. The $I$ band light curves (normalized with respect to their mean magnitudes) of the \textit{gold sample} BL~Her modeled-observed pairs using the OGLE database is presented in Fig.~\ref{fig:LC_gold_ogle}.

\end{document}